\documentclass[12pt,fleqn]{article}
\usepackage{amsmath,amssymb,graphicx,epsfig}
\usepackage[footnotesize,width=0.8\textwidth]{caption}

\def\bea{\begin{eqnarray}}
\def\eea{\end{eqnarray}}
\def\be{\begin{equation}}
\def\ee{\end{equation}}
\def\ba{\begin{array}}
\def\ea{\end{array}}
\def\nn{\nonumber}
\def\a{& \hspace{-10pt}}
\def\b{& \hspace{-7pt}}
\def\summ{\mbox{\large ${\sum}$}}

\def\s#1{\text{\small $#1$}}
\def\t#1{\text{\tiny $#1$}}

\newcommand{\bal}{\begin{aligned}}
\newcommand{\eal}{\end{aligned}}

\font\tenrsfs=rsfs10
\font\sevenrsfs=rsfs7
\font\fiversfs=rsfs5
\newfam\rsfsfam
\textfont\rsfsfam=\tenrsfs
\scriptfont\rsfsfam=\sevenrsfs
\scriptscriptfont\rsfsfam=\fiversfs
\def\mathscr#1{{\fam\rsfsfam\relax#1}}

\begin{document}

\thispagestyle{empty}

\begin{center}

$\;$

\vspace{1cm}

{\LARGE \bf The lightest scalar in theories with \\[2mm] broken supersymmetry}

\vspace{1.3cm}

{\large {\bf Leonardo~Brizi} and {\bf Claudio~A.~Scrucca}}\\[2mm] 

\vspace{0.4cm}

{\large \em Institut de Th\'eorie des Ph\'enom\`enes Physiques\\ 
Ecole Polytechnique F\'ed\'erale de Lausanne\\ 
CH-1015 Lausanne, Switzerland\\}

\vspace{0.2cm}

\end{center}

\vspace{1cm}

\centerline{\bf \large Abstract}
\begin{quote}

We study the scalar mass matrix of general supersymmetric theories with local gauge symmetries, 
and derive an absolute upper bound on the lightest scalar mass. This bound can be saturated by 
suitably tuning the superpotential, and its positivity therefore represents a necessary and sufficient 
condition for the existence of metastable vacua. It is derived by looking at the subspace of all those 
directions in field space for which an arbitrary supersymmetric mass term is not allowed and scalar 
masses are controlled by supersymmetry-breaking splitting effects. This subspace includes not only 
the direction of supersymmetry breaking, but also the directions of gauge symmetry breaking and the 
lightest scalar is in general a linear combination of fields spanning all these directions.  We present 
explicit results for the simplest case of theories with a single local gauge symmetry. For renormalizable 
gauge theories, the lightest scalar is a combination of the Goldstino partners and its square mass is 
always positive. For more general non-linear sigma models, on the other hand, the lightest scalar 
can involve also the Goldstone partner and its square mass is not always positive.

\vspace{5pt}
\end{quote}

\renewcommand{\theequation}{\thesection.\arabic{equation}}

\newpage

\setcounter{page}{1}

\section{Introduction}
\setcounter{equation}{0}

It has been known since the early days of supersymmetry that the spontaneous breaking of 
supersymmetry allows to split the masses of bosons and fermions but not to achieve totally 
arbitrary mass matrices. In general, these mass matrices consist of a supersymmetric contribution 
that is common to all the states of a multiplet plus a non-supersymmetric contribution splitting the
masses of these states within each multiplet. There are then two sources of constraints in such 
mass matrices, which lead to two different kinds of restrictions. 

The first source of constraints is that the various non-supersymmetric contributions to the masses 
are correlated among each other. A simple consequence of these correlations is expressed by 
the celebrated sum rule constraining the supertrace of the full mass matrix. 
When computing this quantity, the supersymmetric contributions to masses drop out and the 
non-supersymmetric contributions combine into a remarkably simple result. This then constrains 
to some extent the relative masses that can be achieved for bosons and fermions, and has 
important implications in phenomenological model building. More precisely, the supertrace of 
the mass matrix vanishes for renormalizable anomaly-free theories \cite{STR1}, whereas it 
depends on the Ricci curvature of the scalar manifold, the derivatives of the gauge kinetic 
function and the trace of the gauge symmetry generators in more general non-linear sigma 
models \cite{STR2}. Similar results also hold true in supergravity theories. Finally, in theories 
with extended supersymmetry these results become even stronger. For instance, in theories 
with rigid N=2 supersymmetry, the supertrace of the mass matrix always vanishes \cite{HKLR}.

The second source of constraints is that some of the supersymmetric contributions to masses 
are fixed by symmetry arguments, and cannot be freely chosen by adjusting the superpotential. 
Most importantly, the supersymmetric contribution to the mass of the Goldstino chiral multiplet 
must vanish, since the fermion of this multiplet is constrained by Goldstone's theorem to have 
vanishing mass. As a result, the two scalar partners of this fermion have masses that are entirely 
controlled by splitting effects. Similarly, the supersymmetric contribution to the mass of the vector 
multiplets is fixed by the values of the gauge symmetry transformations, since the vector boson 
masses arise through the Higgs mechanism. As a result, the real scalar partner of each massive
gauge boson has a mass that differs from the gauge boson mass only by splitting effects, and this 
can also be viewed as the statement that the would-be Goldstone chiral multiplet has a constrained 
mass in the supersymmetric limit. A simple consequence of these restrictions is that there exists 
an upper bound to the mass of the lightest scalar, even if the superpotential is freely tuned. 
The case of theories with only chiral multiplets and no gauge symmetries is well understood.
What matters in this case is the two-dimensional sub-block of the scalar mass matrix restricted 
to the two Goldstino partners. For renormalizable models, the two eigenvalues of this matrix are 
equal and opposite, and the best situation that can occur is that both vanish. This implies the 
presence of two pseudo-moduli fields with vanishing mass, which actually represent flat directions 
of the classical potential with peculiar properties \cite{R,KS}. For more general non-renormalizable 
chiral non-linear sigma-models, one similarly finds that the two eigenvalues are split around an 
average value that is fixed by the Riemann curvature of the K\"ahler manifold, and in the best situation 
one has two scalars with identical masses given by this value \cite{GRS1,GRS2}. Similar results also 
hold in supergravity theories, and these give a useful guideline towards the ingredients that 
are needed to achieve metastable de Sitter vacua in string models \cite{DD, CGGLPS, CGGPS}. The 
case of theories involving also vector multiplets and local gauge symmetries is more complicated and 
less understood (see for example \cite{DF,DT,M} for some simple examples). 
In this case, one should in principle look at a higher-dimensional sub-block of the 
scalar mass matrix that includes not only the two Goldstino partners but also the Goldstone partners.
It has been argued in \cite{GRS3} that the presence of $D$-type in 
addition to $F$-type supersymmetry breaking tends to improve the situation, at least as far as the 
masses of the two Goldstino partners are concerned. But a full analysis including also the Goldstone 
partners is still missing. Finally, in theories with extended supersymmetry, similar but even stronger results 
hold true. For instance, in theories with rigid N=2 supersymmetry some of the Goldstino partners are unavoidably 
tachyonic or at best massless in all the situations where supersymmetry breaking is of $F$ type from the 
N=1 viewpoint, namely models involving only hyper multiplets \cite{GRLS} or only Abelian vector 
multiplets \cite{CKVDFDG}. On the other hand, it has been argued in \cite{JS} that such tachyonic Goldstino 
partners can be avoided in more general situations where supersymmetry breaking is also of $D$ type 
from the N=1 viewpoint, like in particular models involving non-Abelian vector multiplets or charged 
hyper multiplets. But a general study of the masses of the potentially equally dangerous Goldstone partners 
is again missing, although some explicit supergravity examples have been studied in detail \cite{FTV,O,RR}. 
In this same context, it has also been shown in \cite{AB} that under certain assumptions there 
exists an algebraic obstruction against a consistent non-linear realization of N=2 supersymmetry, and 
it would be interesting to assess whether this captures the same information as the presence of tachyons.

The purpose of this paper is to perform a detailed study of the scalar mass matrix of generic theories with 
rigid N=1 supersymmetry and local gauge symmetries, and to derive an upper bound on the value of its 
lightest eigenvalue. The main improvement that we aim to achieve compared to previous analyses 
is to obtain the strongest possible bound, with the property that it should be possible to saturate it by adjusting 
only the superpotential. To achieve this goal, we will need not only to consider the effect of the vector multiplets 
on the two Goldstino partners, but also to include in the analysis the Goldstone partners, and focus our attention 
on the full dangerous sub-block of the scalar mass matrix for which supersymmetric effects are constrained. 

The main result that we will derive in this work is that the most dangerous scalar field is in general a 
linear combination of both the Goldstino and the Goldstone partners. We will moreover argue that the 
maximal value that the mass of this mode can take provides the universal upper bound on the 
scalar masses of the theory that we are looking for, with the property that it can be saturated by 
tuning the superpotential. 
In the simplest case where there is a single spontaneously broken gauge symmetry, we will be 
able to obtain a quite explicit expression for this upper bound. More precisely, denoting by 
$f^i$ and $x^i$ the orthonormal vectors defining the Goldstino and the Goldstone directions 
in field space, and with $m^2_{f \bar f}$, $m^2_{x \bar x}$ and $m^2_{f \bar x}$ the matrix 
elements of the Hermitian block $m^2_{i \bar \jmath}$ of the scalar mass matrix along these 
directions, this bound will be shown to be given by:
\bea
m^2 = {\rm max}
\bigg\{\s{\frac 12} \big(m^2_{f \bar f} + 2\, m^2_{x \bar x}\big) 
- \s{\frac 12} \sqrt{\big(m^2_{f \bar f} - 2\, m^2_{x \bar x}\big)^2\! + 8\, |m^2_{f \bar x}|{\raisebox{10pt}{$$}}^2} \bigg\}\,.
\label{boundanticipated}
\eea
This strongest bound is always smaller-or-equal than the weaker bound ${\rm max} \big\{m^2_{f \bar f}\big\}$ 
that can be derived by looking only at the Goldstino direction, independently of the optimization over the choice 
of the vacuum point and the directions $f^i$ and $x^i$ that defines these bounds. In the particular case of renormalizable 
theories, the optimal choice can be clearly identified and is seen to correspond to a maximization of the value
of the $D$ auxiliary field of the involved vector multiplet. The bound then takes the very explicit form 
$m^2 = |q_{\rm max}/q_{\rm min}| M^2$, where $M$ is the mass of the gauge field
whereas $q_{\rm min}$ and $q_{\rm max}$ denote the smallest and largest charges with common sign. 
In the more general case of non-renormalizable theories, the optimal choice depends also on the curvature 
of the scalar manifold and not just on the structure of the gauging, and can no longer be explicitly determined. 
It is then not possible to make the bound more explicit without specializing to a particular model.

The rest of the paper is organized as follows. In section 2 we review the general structure of 
supersymmetric theories with gauge symmetries. In section 3 we describe the form of 
the scalar mass matrix and study its restriction to the subspace defined by the Goldstino and Goldstone 
directions. In section 4 we derive a general upper bound on the lightest scalar mass,
focusing on theories with a single spontaneously broken gauge symmetry where the relevant matrix 
is three-dimensional and can be studied analytically. In section 5 we study the special case of renormalizable 
theories and show that in that case the lightest scalar in the optimal situation is always a combination of 
just the Goldstino partners,  with a positive square mass depending on the charges. In section 6 we discuss 
the qualitative features of the more general case of non-renormalizable theories and argue that in that case 
the lightest scalar in the optimal situation is a combination of the partners of not only the Goldstino but also 
the Goldstone modes, with a square mass of indefinite sign that depends both on the curvature and the 
structure of the gauging. In section 7 we present our general conclusions. Finally, in appendix A we study 
in some detail a few concrete examples of models to illustrate our general results.

\section{General supersymmetric theories}
\setcounter{equation}{0}

Let us consider a generic $N=1$ theory with $n$ chiral multiplets $\Phi^i$ and $k$ vector multiplets $V^a$. 
The most general two-derivative Lagrangian for such a theory is specified by a real K\"ahler potential 
$K$, a holomorphic superpotential $W$, a holomorphic gauge kinetic function $H_{ab}$ and some 
holomorphic Killing vectors $X_a^i$ generating a group of isometries:
\bea
\a\a \mathcal{L} = \int \! d^4 \theta\, K(\Phi,\bar \Phi,V)
+ \int \! d^2 \theta\, \Big[W(\Phi) + \s{\frac 14} H_{ab} (\Phi) \,W^{a \alpha} W_\alpha^b \Big] + {\rm h.c.} \,.
\eea
In view of taking the Wess-Zumino gauge, we can study this theory by expanding in powers of 
$V^a$. From now on, we will then denote by $K$ the K\"ahler potential at vanishing $V^a$. This defines 
a metric $g_{i \bar \jmath} = K_{i \bar \jmath}$, a Christoffel symbol $\Gamma^k_{ij} = g^{k \bar l} K_{i j \bar l}$ 
and a Riemann tensor $R_{i \bar \jmath k \bar l} = K_{i \bar \jmath k \bar l} - g^{p \bar q} K_{i k \bar q}K_{\bar \jmath \bar l p}$ 
for the scalar field geometry. The usual coordinate-covariant derivatives for this geometry will be~denoted~by~$\nabla_i$.
For later convenience, we shall furthermore introduce an arbitrary gauge coupling constant $g$, although this could be reabsorbed 
in the normalization of $H_{ab}$. The gauge transformations then act as $\delta \Phi^i = \Lambda^a X_a^i$ on the chiral multiplets 
and as $\delta V^a = - \frac i2 g^{\text{--}1}(\Lambda^a - \bar \Lambda^a) + \frac 12 f_{bc}{}^a (\Lambda^b + \bar \Lambda^b) V^c 
+ {\cal O} (V^2)$ on the vector multiplets. The former correspond to general non-linear transformations 
of the scalar fields involving the functions $X_a^i$ and linear transformations of the fermion fields 
involving the scalar-dependent matrices $Q_{a}{}^i{}_j = i \nabla_j X_{a}^i$, whereas the latter correspond 
to the usual transformations of the gauge fields and gaugini involving the structure constants $f_{ab}{}^c$.
We exclude for simplicity the possibility of non-zero variations that amount to a non-trivial K\"ahler transformation, 
since such situations are not guaranteed to be compatible with a coupling to gravity and can also not emerge in 
low-energy effective descriptions of microscopic theories where the variations were strictly vanishing (see \cite{DKS} 
for a recent discussion of this point). In particular, we thus exclude Fayet-Iliopoulos terms. The gauge invariance of the 
Lagrangian then implies the following conditions:
\bea
\a\a X_a^i K_i = \s{\frac i2} g^{\text{--}1} K_a \,, \label{Kinv} \\[-0.5mm]
\a\a X_a^i W_i = 0 \,, \label{Winv} \\[0.5mm]
\a\a X_a^i H_{bci} = - 2 f_{a\t{(}b}^{\;\;\;\;d} H_{c\t{)}d} \label{finv} \,.
\eea
In addition to these conditions, one also has to impose the equivariance condition on the Killing vectors:
\bea
\a\a g_{i \bar \jmath} X_{\t{[}a}^i \bar X_{b\t{]}}^{\bar \jmath} = \s{\frac i4} g^{\text{--}1} f_{ab}^{\;\;\; c} K_c 
\label{equiv}\,.
\eea
The derivative of (\ref{Kinv}) implies that $K_{ai} = 2 i g \bar X_{ai}$, which shows that $-\frac 12 g^{\text{--}1}K_a$ 
can be identified with the Killing potentials for the Killing vectors $X_a^i$. Moreover, eq.~(\ref{equiv}) guarantees 
that these can be chosen to transform in the adjoint representation. 

In the following, we shall for simplicity restrict to the special case where the gauge kinetic function 
$H_{ab}$ is constant, so that $H_{abi} = 0$. This does not represent a very big conceptual limitation, 
but it leads to a substantial simplification of the theory. The condition (\ref{finv}) then states that 
the structure constants with the upper index lowered with the gauge kinetic function should be totally
antisymmetric. This implies that $H_{ab}$ should be equal to some constant real matrix $h_{ab}$ 
proportional to the Killing metric of the gauge group, which in most of the cases is just the identity 
matrix. We shall then assume that 
\bea
\a\a H_{ab} = h_{ab} \,,\;\; H_{abi} = 0 \,.
\eea
We shall on the other hand retain the possibility of having a generic K\"ahler potential $K$ and 
generic Killing vectors $X_a^i$ defining non-constant $Q_a{}^i{}_j$. 
The particular case of renormalizable gauge theories corresponds to choosing 
$K = \delta_{ij} \Phi^i \bar \Phi^{\bar \jmath}$, $X_a^i = - i\, T_a{}^i{}_j \Phi^j$ 
and $Q_a{}^i{}_j = T_a{}^i{}_j$, with constant $T^a{}^i{}_j$.

In the Wess-Zumino gauge, the Lagrangian for the physical component fields $\phi^i$, $\psi^i$,
$A_\mu^a$ and $\lambda^a$ is given by the following expression:
\bea
\a\a {\cal L} = - g_{i \bar \jmath} \, D_\mu \phi^i D^\mu \bar \phi^{\bar \jmath}
-\,i g_{i \bar \jmath} \, \psi^i \big(D\!\!\!\!/\,\bar \psi^{\bar \jmath} + 
\Gamma^{\bar \jmath}_{\bar m \bar n} \, D\!\!\!\!/\, \bar \phi^{\bar m} \bar \psi^{\bar n} \big) 
- \s{\frac 14} h_{ab} \, F^a_{\mu\nu} F^{b\mu\nu} \nn \\
\a\a \hspace{24pt} - \s{\frac i2} h_{ab} \, \lambda^{a} D\!\!\!\!/\, \bar \lambda^b + {\rm h.c.}
+ g^{i \bar \jmath} \, W_i \bar W_{\bar \jmath} + \s{\frac 18} \, h^{ab} K_a K_b 
+ \s{\frac 12} \nabla_i W_j \, \psi^i \psi^j + {\rm h.c.} \nn \\
\a\a \hspace{24pt}  +\, \s{\sqrt{2}}\, g\, g_{i \bar \jmath} \, \bar X_a^{\bar \jmath} \, \psi^i \lambda^a + {\rm h.c.} 
- \s{\frac 14} R_{i \bar \jmath k \bar l} \, \psi^i \psi^k \bar \psi^{\bar \jmath} \bar \psi^{\bar l} \,.
\eea
In the above expression, $F^a_{\mu \nu} = \partial_\mu A^a_\nu - \partial_\nu A_\mu^a + g f_{bc}^{\;\;\;a} A_\mu^b A_\nu^c$ 
is the gauge field-strength and 
$D_\mu \phi^i = \partial_\mu \phi^i + g A^a_\mu X_a^i$, $D_\mu \psi^i = \partial_\mu \psi^i - i g A^a_\mu \,Q_a{}^i{}_j \,\psi^j$,
$D_\mu \lambda^a = \partial_\mu \lambda^a + g f_{bc}^{\;\;\;a} A_\mu^b \lambda^c$ are the gauge-covariant derivatives.

The vacuum is defined by constant values of the scalars $\phi^i$ and vanishing values of the fermions $\psi^i,\lambda^a$ 
and the vectors $A_\mu^a$, minimizing the energy. The values of the auxiliary fields $F^i$ and $D^a$ are then fixed by 
their equations of motion and read:
\bea
\a\a F^i = - g^{i \bar \jmath} \bar W_{\bar \jmath} \,,\;\;
D^a = - \s{\frac 12} h^{ab} K_b = i g h^{ab} X_b^i K_i = - i g h^{ab} \bar X_b^{\bar \imath} K_{\bar \imath}\,.
\eea
The vacuum energy $V$ is given by 
\bea
\a\a V = g_{i \bar \jmath} F^i \bar F^{\bar \jmath} + \s{\frac 12} h_{ab} D^a D^b \,.
\eea
The stationarity condition $V_i = 0$ implies that 
\bea
\a\a \nabla_i W_j \,F^j + i g \bar X_{a i} D^a = 0 \,.
\label{stat}
\eea
Finally, one may easily compute the masses for the modes describing small fluctuations around such a vacuum. 
The scalar square masses are given by
\bea
m^2_{i \bar \jmath} \b=\b g^{k \bar l} \nabla_i W_k \nabla_{\bar \jmath} \bar W_{\bar l} - R_{i \bar \jmath k \bar l} \, F^k \bar F^{\bar l} 
+ g^2 h^{ab} \bar X_{a i} X_{b \bar \jmath} + g\, Q_{a i \bar \jmath} D^a \,, 
\label{m0vec}\\[0mm]
m^2_{ij} \b=\b - \nabla_i \nabla_j W_K \, F^K - g^2 h^{ab} \bar X_{a i} \bar X_{b j} \,.
\eea
The fermion masses are instead found to be
\bea
\mu_{ij} = \nabla_i W_j \,,\;\; \mu_{ab} = 0 \,,\;\; \mu_{ia} = \s{\sqrt{2}}\, \bar X_{ai} \,.
\eea
Finally, the vector boson square masses read
\bea
\a\a M^2_{ab} = 2 g^2 g_{i \bar \jmath} X_{\t{(}a}^i \bar X_{b\t{)}}^{\bar \jmath} \,.
\eea
The vacuum is at least metastable if the full mass matrix for the scalar fluctuations turns out to be a 
positive definite matrix.

The global supersymmetry is spontaneously broken whenever $V \neq 0$, that is whenever some of the auxiliary 
fields $F^i$ or $D^a$ take non-vanishing values. In that case there exists a physical Goldstino fermion 
$\eta \propto \bar F_i \psi^i + \frac i{\sqrt{2}} \, D_a \lambda^a$ with vanishing mass $\mu_\eta = 0$. 
Notice however that by contracting the stationarity condition (\ref{stat}) 
with the Killing vectors $X_a^i$, taking the imaginary part and finally using (\ref{Winv}) as well as its derivative, one finds 
the following relation between the values of $F^i$ and $D^a$ at stationary points:
\bea
\a\a Q_{a i \bar \jmath} F^i \bar F^{\bar \jmath} - \s{\frac 12} g^{\text{--}1}M^2_{ab} D^b = 0 \,. 
\label{relDFF}
\eea
Similarly, by contracting (\ref{stat}) with the auxiliary fields $F^i$, one deduces that:
\bea
\a\a \mu_{ij} F^i F^j = 0 \,.
\eea
These expressions show that the basic source of supersymmetry breaking must come from the chiral auxiliary 
fields $F^i$, whereas the vector auxiliary fields $D^a$ can only give additional effects whose sizes are linked 
to the masses of the vector bosons. 

The local gauge symmetries are spontaneously broken whenever $M^2_{ab} \neq 0$. In that case there exist
unphysical would-be Goldstone scalars $\sigma_a \propto \bar X_{ai} \phi^i + X_{a\bar \imath} \bar \phi^{\bar \imath}$
with formally vanishing masses $m_{\sigma_a} = 0$. But these modes are in fact absorbed by the gauge bosons 
through the Higgs mechanism, and therefore map to physical degrees of freedom that are massive.
In the same process, the combinations of chiral fermions $\chi_a \propto \bar X_{ai} \psi^i$ pair with the gaugini 
$\lambda^a$ to give massive Dirac fermions.

The mass spectrum displays a rather intricate structure in the general situation in which both supersymmetry 
and the gauge symmetries are broken. As discussed above, the relevant complex directions defining these two 
breakings are respectively $F^i$ and $X_a^i$, and gauge invariance of the superpotential implies that these 
are orthogonal to each other: $g_{i \bar \jmath} F^i \Bar X_a^{\bar \jmath} = 0$. When supersymmetry is unbroken, 
the situation simplifies and can be understood in terms of multiplets. The $k$ chiral multiplets corresponding to the 
directions $X_a^i$ are generically absorbed by the $k$ vector multiplets through a supersymmetric Higgs mechanism. 
In a super-unitary gauge, one is then left with $n - k$ chiral multiplets corresponding to the directions orthogonal 
to $X_a^i$ plus $k$ unconstrained vector multiplets. The physical square-mass spectrum then consists of $n-k$ levels 
corresponding to the eigenvalues of the matrix $g^{k \bar l} W_{ik} \bar W_{\bar \jmath \bar l}$ restricted to the subspace 
orthogonal to the $X_a^i$, each containing two real scalars and one two-component fermion, and $k$ levels 
corresponding to the eigenvalues of the matrix $2 g^2 g_{i \bar \jmath} X_{\t{(}a}^i \bar X_{b\t{)}}^{\bar \jmath}$, each
containing one real scalar, two two-component fermions and one three-component vector.
When supersymmetry is broken, on the other hand, additional mass splittings are generated with respect to the above 
spectrum, and the situation becomes more complicated. But the essential modification with respect to the previous case
is rather simple. In each chiral multiplet the two real scalars can split from the fermion and in each unconstrained vector multiplet 
the real scalar can split from the fermions and the vector. In addition, one linear combination of all the fermions must be 
exactly massless.

A well-known general result about the above mass matrices that holds true at any point, even if supersymmetry 
and the gauge symmetries are both broken, is the supertrace of the full square-mass matrix. This is found to be 
given by ${\rm str}\, {\cal M}^2 = 2\, R_{i \bar \jmath}\, F^i \bar F^{\bar \jmath} + 2g\, Q_a{}^i{}_i \, D^a$.
At a generic stationary point, one may further simplify this result by using the relation 
(\ref{relDFF}). In this way, one finally finds:
\bea
\a\a {\rm str}\, {\cal M}^2 = 2 \Big[R_{i \bar \jmath} 
+ 2\, g^2 \, Q_a{}^k{}_k \, M^{\text{--}2ab} \, Q_{b i \bar \jmath} \Big] F^i \bar F^{\bar \jmath} \,.
\eea
The value taken by the right-hand side restricts to some extent the relative values that 
bosons and fermion masses can take. 

\section{Structure of the scalar mass matrix}
\setcounter{equation}{0}

Let us now study more specifically the masses of scalar fields. Since the two real components of 
each complex scalar field are allowed to split, one has to consider the space of all the independent
real modes. This can be described by $2n$-dimensional vectors $\Phi^I$ built out of the $n$ fields 
$\phi^i$ and their complex conjugates $\bar \phi^{\bar \imath}$: 
\bea
\a\a \phi^I = \big(\phi^i \;\; \bar \phi^{\bar \imath} \big) \,,\;\;
\phi^{\bar J} = 
\left(
\begin{matrix} 
\bar \phi^{\bar \jmath} \smallskip \\
\phi^j \\
\end{matrix}
\right) \,.
\eea
With this parametrization, the quadratic Lagrangian for the scalar fields can be written in the following form:
\bea
\a\a {\cal L} = \s{\frac 12} g_{I \bar J} \partial_\mu \phi^I \partial^\mu \phi^{\bar J} - \s{\frac 12} m^2_{I \bar J} \phi^I \bar \phi^{\bar J} \,,
\eea
with wave-function and square-mass matrices given by
\bea
g_{I \bar J} = 
\left(
\begin{matrix} 
g_{i \bar \jmath} \!&\! 0 \smallskip\ \\
0 \!&\! g_{\bar \imath j}  \\
\end{matrix}
\right) \,,\;\;
m^2_{I \bar J} = 
\left(
\begin{matrix} 
m^2_{i \bar \jmath} \!&\! m^2_{i j} \smallskip\ \\
m^2_{\bar \imath \bar \jmath} \!&\! m^2_{\bar \imath j}  \\
\end{matrix}
\right) \,.
\eea
To obtain the physical masses, one can then proceed as follows. First, one choses a parametrization of the fields 
such that the wave-function $g_{I \bar J}$ locally trivializes to the identity matrix and the kinetic terms are canonically 
normalized. This corresponds to choosing normal coordinates around the vacuum point. Next, one diagonalizes 
the Hermitian matrix $m^2_{I \bar J}$ to find the mass eigenvalues $m^2_{(I)}$. Equivalently, one can consider
the matrix $m^2_{I \bar J}$ in a new basis defined by a set of vectors $v_{K}^I$ that are orthonormal with respect 
to the metric $g_{I \bar J}$. The eigenvalues of the new matrix defined by all the matrix elements of $m^2_{I \bar J}$ 
on the basis of vectors $v_K^I$ then yield directly the physical masses. This is the approach that we will use.

To make progress in our quest for an interesting bound on the physical mass eigenvalues, and in particular the 
minimal physical eigenvalue $m^2_{\rm min}$, we will use some standard results in linear algebra. The basic point is that the 
value of the matrix $m^2_{I \bar J}$ along any particular direction must be larger that $m^2_{\rm min}$. A slight 
generalization of this is that the eigenvalues of any sub-block of the matrix $m^2_{I \bar J}$, corresponding for 
example to the subspace spanned by a set of several particular directions, must similarly be all larger than 
$m^2_{\rm min}$. This means that we can find an upper bound to $m^2_{\rm min}$ by computing the smallest 
eigenvalue of any principal sub-matrix of $m^2_{I \bar J}$. In general, the obtained bound improves in quality 
by considering larger and larger sub-matrices, and the exact value of $m^2_{\rm min}$ can be obtained only 
by considering the full matrix. Nevertheless, there is a well-defined limiting situation in which the bound derived 
by considering a finite diagonal block actually saturates $m^2_{\rm min}$. This happens when the complementary 
diagonal block has eigenvalues that are very large compared to the elements of the off-diagonal block. 
For this reason, to detect the obstructions against making $m^2_{\rm min}$ large it is enough to study the mass 
matrix along those directions where its values cannot be made arbitrarily large by adjusting the superpotential. 

Each direction defined by a unit vector $v^i$ in the space of complex scalar fields $\phi^i$ defines a 
plane in the space of real scalar fields $\phi^I$, which can be described by a basis of two orthonormal unit 
vectors $v^I_+$ and $v^I_-$ defined as follows:
\bea
\a\a v^I_+ = \s{\frac 1{\sqrt{2}}} \big(v^i \;\; \bar v^{\bar \imath} \big) \,,\;\;
v^I_- =  \s{\frac 1{\sqrt{2}}} \big(i v^i \;\, \text{--}\, i \bar v^{\bar \imath} \big) \,.
\eea
Strictly speaking, the vector space of all real scalar fields is a real vector space, and one is therefore 
allowed to perform only real orthogonal transformations. However, for the problem of studying the 
eigenvalues of the mass matrix $m^2_{I \bar J}$, which is Hermitian, one may also consider complex 
unitary transformations, because such more general transformations still preserve these eigenvalues. 
For a given complex direction $v^i$, one may then also use as alternative basis the two orthonormal 
vectors $v_{A,B} = \frac 1{\sqrt{2}} (v_+^I \mp i v_-^I)$, which take the form:
\bea
\a\a v^I_A = \big(v^i \;\; 0 \big) \;,\;\;
v^I_B = \big(0 \;\; \bar v^{\bar \imath} \big) \,.
\eea

From the discussion of previous section, we know that there are two kinds of special complex directions
along which the mass matrix displays particular restrictions. These are the supersymmetry-breaking Goldstino 
direction $F^i$ and the gauge-symmetry-breaking Goldstone directions $X_a^i$. In all the other orthogonal 
directions, one can have arbitrary supersymmetric contributions to the mass. Taking these to be large one can 
then forget about these extra directions altogether, as already explained. Let us then focus on the subspace defined 
by the complex directions $F^i$ and $X_a^i$. We already know that $F^i$ is always orthogonal to all the 
$X_a^i$, as a consequence of the gauge invariance of the superpotential. On the other hand, the $X_a^i$ 
are in general not orthogonal to each other, and the matrix of their scalar products defines in fact the vector 
mass matrix. We may however perform an orthogonal transformation in the space of vector multiplets, to 
go to a basis where at the vacuum all the $X_a^i$ are orthogonal to each other and the vector mass 
matrix is diagonal. The norms of the vectors $F^i$ and $X_a^i$ define respectively the supersymmetry 
breaking scale $\sqrt{|F|}$ in the chiral multiplet sector and the masses $M_a$ of the vector fields.
More precisely, these quantities are defined as follows:
\bea
\a\a |F| = \raisebox{0pt}{$\sqrt{g_{i \bar \jmath} F^i \bar F^{\bar \jmath}}$} \,,\;\; 
M_a = \s{\sqrt{2}} g \raisebox{0pt}{$\sqrt{g_{i \bar \jmath} X_a^i \bar X_a^{}{\!\!}^{\bar \jmath}}$}  \,.
\eea
One then finds:
\bea
\a\a g_{i \bar \jmath} F^i \bar F^{\bar \jmath} = |F|^2 \,,\;\;
g_{i \bar \jmath} X_a^i \bar X_b^{\bar \jmath} = \s{\frac 12} g^{\text{--}2}M_a M_b \, \delta_{ab} \,,\;\;
g_{i \bar \jmath} F^i \bar X_b^{\bar \jmath} = 0 \,.
\eea
We may finally define the following normalized vectors:
\bea
\a\a f^i = \frac {F^i}{\sqrt{F^k \bar F_k}} = \frac {F^i}{F}\,,\;\;
x_a^i = \frac {X_a^i}{\sqrt{X_a^k \bar X_{ak}}} = \s{\sqrt{2}} g \frac {X_a^i}{M_{a}}\,.
\eea
These form an orthonormal basis for the subspace of complex directions we 
want to study, and satisfy:
\bea
\a\a g_{i \bar \jmath} \, f^i \bar f^{\bar \jmath} = 1 \,,\;\; 
g_{i \bar \jmath} \, x_a^i \bar x_b^{\bar \jmath} = \delta_{ab} \,,\;\; 
g_{i \bar \jmath} \, f^i \bar x_b^{\bar \jmath} = 0 \,.
\eea

Following our general discussion on the map between a complex direction in the space 
of complex scalars and a basis of two independent directions in the space of real scalars, 
we now introduce the following orthonormal basis of real directions:
\bea
\a\a f^I_+ =  \s{\frac 1{\sqrt{2}}} \big(f^i \;\; \bar f^{\bar \imath} \big) \,,\;\;
f^I_- =  \s{\frac 1{\sqrt{2}}} \big(i f^i \;\, \text{--}\, i \bar f^{\bar \imath} \big) \,, \\
\a\a x^I_{a+} = \s{\frac 1{\sqrt{2}}} \big(x_a^i \;\; \bar x_a^{\bar \imath} \big) \,,\;\;
x^I_{a-} =  \s{\frac 1{\sqrt{2}}} \big(i x_a^i \;\, \text{--}\, i \bar x_a^{\bar \imath} \big) \,.
\eea
Alternatively, we may as already explained also use the alternative but less physical basis 
defined by
\bea
\a\a f^I_A = \big(f^i \;\; 0 \big) \,,\;\;
f^I_B = \big(0 \;\; \bar f^{\bar \imath} \big) \,, \\
\a\a x^I_{aA} = \big(x_a^i \;\; 0 \big) \,,\;\;
x^I_{aB} = \big(0 \;\; \bar x_a^{\bar \imath} \big) \,.
\eea
The directions $f^I_+$ and $f^I_-$ describe the two real scalar partners of the massless Goldstino fermion. 
Due to the symmetric roles of these two modes, it will in fact be convenient to use the alternative description in 
terms of $f^I_A$ and $f^I_B$. In the limit of unbroken supersymmetry, the modes defined by $f^I_+$ and $f^I_-$ 
would both belong to the same multiplet as the massless Goldstino fermion and would thus be massless too. 
As a result, their masses can be non-zero only because of splitting effects. 
The directions $x^I_{a+}$ and $x^I_{a-}$ describe instead two different kinds of real scalars which are respectively 
the unphysical would-be Goldstone modes, which correspond to fake null vectors of the mass matrix that we should 
discard, and their partners, which we should instead consider. Due to the asymmetric
roles of these two kinds of modes, it will not be convenient to use the alternative description in terms of 
$x^I_{aA}$ and $x^I_{aB}$. In the limit of unbroken supersymmetry, the modes $x^I_{a-}$ would belong to 
the same multiplet as the massive vector bosons and would thus be massive too. As a result, their mass 
can differ from that of the gauge fields only by splitting effects. We thus find 
a total of $2 + k$ scalar modes which are dangerous for metastability: the $2$ modes associated to $f^I_\pm$ 
and alternatively described by $f^I_{A,B}$, whose masses are equal to zero plus supersymmetry breaking effects, 
and the $k$ modes associated to $x_{a-}^I$, whose masses are equal to the gauge boson masses plus 
supersymmetry breaking effects. 

Let us then look at the mass matrix $m^2_{I \bar J}$ in the $(2 + k)$-dimensional subspace spanned by the 
vectors $f^I_A = (f^i \;\; 0)$, $f^I_B = (0 \;\; \bar f^{\bar \imath})$ and $x^I_{a-} = (i x_a^I \;\, \text{--}\, i \bar x_a^{\bar \imath})$,
which form an orthonormal set. More precisely, we need to compute the matrix elements 
$m^2_{\alpha \bar \beta} = m^2_{I \bar J} v_\alpha^I \bar v_{\bar \beta}^{\bar J}$,
where $v_\alpha^I$ can be either $f^I_A$, $f^I_B$ or $x^I_{a-}$.
Exploiting gauge invariance, we can rewrite most of the contributions coming from the non-Hermitian 
blocks $m^2_{ij}$ and $m^2_{\bar \imath \bar \jmath}$ in terms of the Hermitian blocks $m^2_{i \bar \jmath}$. 
Indeed, Goldstone's theorem implies that $m^2_{i j} x_a^j = - m^2_{i \bar \jmath} \bar x_a^{\bar \jmath}$ at a 
stationary point. One then finds that the $(2 + m)$-dimensional sub-matrix $m^2_{\alpha \bar \beta}$ takes 
the form
\bea
\a\a m^2_{\alpha \bar \beta} = 
\left(
\begin{matrix} 
m^2_{f \bar f} \!&\! \Delta \!&\! \text{--} \s{\sqrt{2}} i\, m^{2*}_{f \bar x_b} \smallskip\ \\
\Delta^* \!&\! m^2_{f \bar f} \!&\! \s{\sqrt{2}} i\, m^{2}_{f \bar x_b} \smallskip\ \\
\s{\sqrt{2}} i\, m^{2}_{f \bar x_a} \!&\! \text{--} \s{\sqrt{2}} i\, m^{2*}_{f \bar x_a} \!&\! 2\, m^2_{x_a \bar x_b}
\end{matrix}
\right) \,, \label{mat}
\eea
where
\bea
\a\a m^2_{f \bar f} = m^2_{i \bar \jmath} \, f^i \bar f^{\bar \jmath} \,,\;\;
m^2_{f \bar x_b} = m^2_{i \bar \jmath} \, f^i \bar x_b^{\bar \jmath} \,,\;\;
m^2_{x_a \bar x_b} = m^2_{i \bar \jmath} \, x_a^i \bar x_b^{\bar \jmath} \,,
\eea
and
\bea
\a\a \Delta = m^2_{i j} f^i f^j \,.
\eea
It is important to emphasize that the above structure is completely general, since it depends 
only on the gauge invariance of the theory and not on the detailed structure of the masses. 

It is a straightforward exercise to compute the entries $m^2_{f \bar f}$, $m^2_{f \bar x_b}$ and $m^2_{x_a \bar x_b}$,
which are given by the Hermitian block $m^2_{i \bar \jmath}$ of eq.~(\ref{m0vec}) along the directions defined by 
$F^i$ and $X_a^i$. The resulting expressions can be significantly simplified by making use of the stationarity condition, 
which holds at the vacuum, as well as the 
relations implied by gauge invariance, which hold at any point and can therefore also be differentiated. Most importantly, 
the dependence on the second derivatives of the superpotential can be completely eliminated. Defining the obvious notation
$R_{v \bar w y \bar z} = R_{i \bar \jmath k \bar l} \, v^i \bar w^{\bar \jmath} y^k \bar z^{\bar l}$ and 
$Q_{a\hspace{1pt} v \bar w} = Q_{i \bar \jmath} \, v^i \bar w^{\bar \jmath}$ 
for any complex directions $v^i$, $w^i$, $y^i$ and $z^i$,  and recalling that $M^2_{ab} = M_a M_b \, \delta_{ab}$, 
one finds:
\bea
\a\a m^2_{f \bar f} = - \bigg[R_{f \bar f f \bar f} - 4g^2 \, \summ_c \frac {Q_{c f \bar f} \,Q_{c f \bar f}}{M_c^2} \bigg] |F|^2 \,,
\label{mff} \\
\a\a m^2_{x_a \bar x_b} \! = \s{\frac 12} M^2_{ab} \!
- \bigg[R_{f \bar f x_a \bar x_b}\! - 2 g^2 \,\summ_c \frac {Q_{c f \bar f} \,Q_{c\hspace{1pt} x_a \bar x_b}}{M_c^2}
- 2 g^2 \frac {(Q_a \!\cdot Q_b)_{f \bar f}}{M_a M_b} \bigg] |F|^2 \,, 
\label{mxaxb}\\
\a\a m^2_{f \bar x_b} = 
- \bigg[R_{f \bar f f \bar x_b} \! - 4 g^2 \, \summ_c \frac {Q_{c f \bar f} \,Q_{c f \bar x_b} \!}{M_c^2} \bigg] |F|^2 .
\label{mfxb}
\eea
The entry $\Delta$ has instead a more complicated expression, and it is not possible to simplify it in any relevant way 
by using the stationarity and the gauge invariance conditions. Most importantly, the dependence on the third derivatives of the 
superpotential cannot be eliminated, and varying such derivatives allows to vary $\Delta$ over the entire complex plane.
Therefore: 
\bea
\a\a \Delta = \text{generic complex number that can be adjusted by tuning $W_{ijk}$} \,.
\eea

We may now ask what is the upper bound on the smallest eigenvalue of the above matrix $m^2_{\alpha \bar \beta}$ 
when $m^2_{f \bar f}$, $m^2_{f \bar x_b}$ and $m^2_{x_a \bar x_b}$ are held fixed and $\Delta$ is freely varied. 
As already explained, this would also represent an upper bound on the smallest eigenvalue $m^2_{\rm min}$ of the 
full mass matrix $m^2_{I \bar J}$. Unfortunately, this question is still quite complicated for generic theories with arbitrary 
gauge symmetries, where $k$ can be arbitrarily large and it is thus difficult to study the full $(2 +k)$-dimensional 
matrix. The importance of the Goldstone directions $x_a$ with respect to the Goldstino direction depends however 
crucially on the relative size of the vector masses $M_a$ compared to the chiral supersymmetry breaking scale $\sqrt{|F|}$. 
When the $M_a$ are much larger than $\sqrt{|F|}$, the situation simplifies substantially and the heavy vector multiplets 
can in fact be integrated out in a supersymmetric way to define an effective theory for the light chiral multiplets. 
The way in which this can be done has been described in some detail in \cite{BS}. In particular, for the sub-sector 
of scalar fields we are focusing on, we see that all the modes associated to $x^I_{a-}$ are very heavy, and their mixings 
with the modes associated to $f^I_A$ and $f^I_B$ have a negligible effect. The only dangerous light modes are then those 
associated with $f^I_A$ and $f^I_B$, and the largest value for the smallest 
mass is obtained by tuning $\Delta$ to zero. The upper bound $m^2_{\rm min}$ is then given by (\ref{mff}), up to negligible 
effects of order ${\cal O}(|F|^4/M_a^2)$, and the square bracket in (\ref{mff}) can be interpreted as the corrected Riemann 
curvature of the effective theory along the Goldstino direction. When the $M_a$ are instead 
comparable-or-smaller than $\sqrt{|F|}$,  the modes associated to $x^I_{a-}$ are a priori as light and as dangerous as 
the modes associated to $f^I_A$ and $f^I_B$, and the problem acquires its full-fledged complication. It is this situation 
that we would like to study in some detail.

For the sake of clarity, we shall mostly restrict our study to the simplest case of theories with a single $U(1)$ 
gauge symmetry and $k=1$. In this case, it is possible to extract analytically the full information and derive a simple 
necessary and sufficient bound, which can be saturated by adjusting the superpotential. In more complicated 
theories with several gauge symmetries forming a more general group $G$, on the other hand, one may 
get some partial analytic information by studying smaller sub-blocks of dimension one, two and three, and derive 
simple necessary but not sufficient bounds, which can a priori not be saturated by adjusting the superpotential.
In particular, one may look separately at all the possible directions in the generator space and figure out which 
one leads to the strongest bound. An obvious naive guess for a special direction to look at in the space of 
generators is the direction $d^a = D^a/|D|$ defined by the vector auxiliary fields $D^a$. This is also suggested
by the fact that $D^a$ appears together with $F^i$ in the definition of the Goldstino fermion. When looking at the 
special direction $x^I_- = d^a x^I_{a-}$, some partial and interesting simplifications do indeed occur in the expressions 
(\ref{mxaxb}) and (\ref{mfxb}), but since we were not able to reach a really simple and useful result 
by pursuing this direction, we will not comment any further on this, and restrict from now to the basic case 
involving only one symmetry generator.

\section{Bound on the lightest scalar mass}
\setcounter{equation}{0}

Let us now consider the case of theories with a single $U(1)$ gauge symmetry, where the index $a$ 
takes a single value and can therefore be dropped. The matrix (\ref{mat}) is then $3$-dimensional, and 
it turns out that it is possible to study the behavior of its eigenvalues in a fully analytic way. In order to illustrate 
the fact that the study of larger sub-blocks of the mass matrix leads to sharper bounds on the lightest eigenvalue, 
we shall however successively study sub-blocks of dimensions one, two and three.

There are three possible principal blocks of dimension one, which correspond to the diagonal elements, but only two 
of them are independent, namely:
\bea
\a\a m^2_{f \bar f} \,,\;\; 2\,m^2_{x \bar x} \,.
\eea
Both of these values represent upper bounds on $m^2_{\rm min}$. Which one is the smallest and thus leads 
to the strongest bound depends however on the situation. We therefore conclude that a first bound that we 
can write is:
\bea
m^2_{\rm min} \le m^2_{(1)} \,,\;\; 
m^2_{(1)} = {\rm min} \big\{m^2_{f \bar f}, 2\,m^2_{x \bar x} \big\} \,.
\label{bound1}
\eea

There are then three possible principal blocks of dimension two, but again only two of these are independent. The first possibility 
is the upper $2$-dimensional block of (\ref{mat}), with two identical diagonal elements given by $m^2_{f \bar f}$ and off-diagonal
element given by $\Delta$. The two eigenvalues of such a matrix are $m^2_{f \bar f} \pm |\Delta|$. The maximal value for the 
smallest of these is achieved by choosing $\Delta = 0$ and is given by $m^2_{f \bar f}$. This sets an upper bound on 
$m^2_{\rm min}$, but this bound is already contained in the previously derived bound (\ref{bound1}). The second possibility 
is the lower $2$-dimensional block of (\ref{mat}), which is given by
\bea
\left(
\begin{matrix} 
m^2_{f \bar f} \!&\! \s{\sqrt{2}} i\, m^{2}_{f \bar x} \smallskip\ \\
\text{--} \s{\sqrt{2}} i\, m^{2*}_{f \bar x} \!&\! 2\, m^2_{x \bar x}\\
\end{matrix}
\right) \,.
\eea
The eigenvalues of this matrix are easily computed and are given by:
\bea
m^2_\pm = \s{\frac 12} \big(m^2_{f \bar f} + 2\, m^2_{x \bar x}\big) 
\pm \s{\frac 12} \sqrt{\big(m^2_{f \bar f} - 2\, m^2_{x \bar x}\big)^2 \! + 8\, |m^2_{f \bar x}|{\raisebox{10pt}{$$}}^2} \,.
\eea
Both of these eigenvalues set upper bounds on $m^2_{\rm min}$. The smallest one that leads to the strongest bound is 
always the one with the negative sign choice. This leads to a new bound, which is always stronger-or-equal than the 
previous bound (\ref{bound1}) and takes into account the non-trivial level-repulsion effect induced by the off-diagonal 
element $m^2_{f \bar x}$:
\bea
m^2_{\rm min} \le m^2_{(2)} \,,\;\; 
m^2_{(2)} = \s{\frac 12} \big(m^2_{f \bar f} + 2\, m^2_{x \bar x}\big) 
- \s{\frac 12} \sqrt{\big(m^2_{f \bar f} - 2\, m^2_{x \bar x}\big)^2 \! + 8\, |m^2_{f \bar x}|{\raisebox{10pt}{$$}}^2} \,.
\label{bound2}
\eea

Finally, one may try to look at the full block of dimension three, which should in this case yield the full information.
This is given by:
\bea
\left(
\begin{matrix} 
m^2_{f \bar f} \!&\! \Delta \!&\! \text{--} \s{\sqrt{2}} i\, m^{2*}_{f \bar x} \smallskip\ \\
\Delta^* \!&\! m^2_{f \bar f} \!&\! \s{\sqrt{2}} i\, m^{2}_{f \bar x} \smallskip\ \\
\s{\sqrt{2}} i\, m^{2}_{f \bar x} \!&\! \text{--} \s{\sqrt{2}} i\, m^{2*}_{f \bar x} \!&\! 2\, m^2_{x \bar x}
\end{matrix}
\right) \,.
\eea
For generic $\Delta$, the eigenvalues of this matrix are quite complicated, since they are determined by 
the roots of a cubic characteristic polynomial. However, their values for the optimal choice of $\Delta$ 
that maximizes the smallest of them can be determined analytically. To understand this, let us first recall 
that by the anti-crossing theorem of Wigner and von Neumann, one generically needs to tune 
two or three real parameters to force the eigenvalue of a real-symmetric or Hermitian matrix to cross.
In our case, the matrix is Hermitian but due to its very special form it actually behaves like a real-symmetric
one. In fact we know that there actually exists a basis where the matrix simplifies from Hermitian to 
real-symmetric. One can then verify that its eigenvalues always cross at isolated points in the $\Delta$ 
complex plane. Knowing this, it becomes clear that the highest value for the minimal eigenvalue is obtained 
at such a crossing point. But since at that point two eigenvalues become degenerate, the cubic characteristic 
polynomial simplifies and it should be possible to solve the problem analytically. One way to derive 
the desired result is to start from the characteristic equation written after decomposing the two complex 
entries $\Delta$ and $m^2_{f \bar x}$ in the form of a modulus times a phase: 
\bea
\a\a \big(\lambda - m^2_{f \bar f}\big)^2 \big(\lambda - 2\, m^2_{x \bar x} \big) 
- 4\, |m^2_{f \bar x}|{\raisebox{10pt}{$$}}^2 \big(\lambda - m^2_{f \bar f}\big) \nn \\
\a\a \,- \, |\Delta|^2 \big(\lambda - 2\, m^2_{x \bar x} \big) 
+ 4\, |\Delta| |m^2_{f \bar x}|{\raisebox{10pt}{$$}}^2 \cos \big(\arg \Delta - 2 \arg m^2_{f \bar x} \big) = 0 \,.
\label{char}
\eea
Form the form of this equation, it is clear that the optimal choice for the phase of $\Delta$ is the one 
minimizing the last term, in such a way that the cosine is equal to $-1$, that is:
\bea
\a\a \arg \Delta = 2 \arg m^2_{f \bar x} + \pi \,.
\label{argDelta}
\eea
Plugging back this expression into the characteristic equation (\ref{char}), this simplifies to 
$\big(\lambda - m^2_{f \bar f} + |\Delta|\big) \big[\big(\lambda - 2\, m^2_{x \bar x} \big) 
\big(\lambda - m^2_{f \bar f} - |\Delta|\big) - 4\, |m^2_{f \bar x}|{\raisebox{10pt}{$$}}^2 \big] = 0$. 
The three solutions of this cubic equation for $\lambda$ are now easy to find analytically and they
are given by $m^2_{f \bar f} - |\Delta|$ and $\s{\frac 12} \big(m^2_{f \bar f} + 2\, m^2_{x \bar x} + |\Delta| \big) 
\pm \s{\frac 12} \big[\big(m^2_{f \bar f} - 2\, m^2_{x \bar x} + |\Delta|\big)^2\! + 16\, |m^2_{f \bar x}|^2\big]^{1/2}$.
The optimal value for $|\Delta|$, which maximizes the minimal eigenvalue, is obtained when the first 
eigenvalue crosses the smallest of the other two, which is the one with the relative minus sign. This fixes:
\bea
\a\a |\Delta| = \s{\frac 12} \big(m^2_{f \bar f} - 2\, m^2_{x \bar x} \big) 
+ \s{\frac 12} \sqrt{\big(m^2_{f \bar f} - 2\, m^2_{x \bar x}\big)^2\! + 8\, |m^2_{f \bar x}|{\raisebox{10pt}{$$}}^2} \,.
\label{modDelta}
\eea
At the optimal point defined by (\ref{argDelta}) and (\ref{modDelta}), the values of the two degenerate lowest 
eigenvalues and the highest eigenvalues are finally given by:
\bea
m^2_\pm = \s{\frac 12} \big(m^2_{f \bar f} + 2\, m^2_{x \bar x}\big) 
\pm \s{\frac 12} \sqrt{\big(m^2_{f \bar f} - 2\, m^2_{x \bar x}\big)^2\! + 8\, |m^2_{f \bar x}|{\raisebox{10pt}{$$}}^2} \,.
\eea
Both of these eigenvalues give upper bounds on $m^2_{\rm min}$. The smallest one that leads to the strongest 
bound is as before the one with the negative sign choice. This leads to a new bound, which is however seen 
to be identical to the previous bound (\ref{bound2}), showing that the potential level-repulsion effect that is induced
by a generic off-diagonal element $\Delta$ can be trivialized by optimally choosing the value of this element 
through a tuning of the superpotential:
\bea
m^2_{\rm min} \le m^2_{(3)} \,,\;\; 
m^2_{(3)} = \s{\frac 12} \big(m^2_{f \bar f} + 2\, m^2_{x \bar x}\big) 
- \s{\frac 12} \sqrt{\big(m^2_{f \bar f} - 2\, m^2_{x \bar x}\big)^2\! + 8\, |m^2_{f \bar x}|{\raisebox{10pt}{$$}}^2} \,.
\label{bound3}
\eea

Summarizing, we have managed to find explicit expressions for the upper bounds 
$m^2_{(1)}$, $m^2_{(2)}$, $m^2_{(3)}$ on the lightest mass that descend from blocks 
of dimension $1$, $2$, $3$. As expected, these are increasingly strong and satisfy:
\bea
\a\a m^2_{(1)} \ge m^2_{(2)} \ge m^2_{(3)} \,.
\eea 
These bounds hold however for a fixed theory at a fixed vacuum. In particular, they depend on the 
direction $f^i$ and on the vacuum coordinates $\phi^i$, which determine the direction $x^i$ and the 
values of $R_{i \bar \jmath k \bar l}$ and $Q_{i \bar \jmath}$. 
We may then derive a more useful and universal bound by further optimizing the superpotential $W$
to maximize the smallest mass. The strongest version 
of this fully optimized bound, which is our main result, then takes the form
\bea
\a\a m^2_{\rm min} \le m^2 \,,
\eea
where
\bea
m^2 = {\rm max}
\bigg\{\s{\frac 12} \big(m^2_{f \bar f} + 2\, m^2_{x \bar x}\big) 
- \s{\frac 12} \sqrt{\big(m^2_{f \bar f} - 2\, m^2_{x \bar x}\big)^2\! + 8\, |m^2_{f \bar x}|{\raisebox{10pt}{$$}}^2} \bigg\}\,.
\label{bound}
\eea
More precisely, the optimization of $W$ defining (\ref{bound}) can be performed as follows. 
At any given point one can adjust $n-1$ independent complex first derivatives~$W_i$, $n(n-1)/2$ 
independent complex second derivatives $W_{ij}$, and $(n-1)n(n+1)/6$ independent complex third derivatives 
$W_{ijk}$, compatibly with gauge invariance. One may then tune the $n-1$ $W_i$ to adjust the direction $f^i$
and the scale $\sqrt{|F|}$, $n-1$ of the $W_{ij}$ to adjust the values of $n-1$ of the fields $\phi^i$ compatibly with 
the $n-1$ stationary conditions in the non-Goldstone directions,
and finally $1$ of the $W_{ijk}$ to adjust the quantity $\Delta$ to its optimal value. In this optimized situation, 
however, there is still $1$ combination of fields $\phi^i$ related to the vector mass $M^2 = 2\, g^2 |X|^2$ that
cannot be freely adjusted, because the stationarity condition (\ref{relDFF}) along the Goldstone direction does not 
depend on $W_{ij}$ and $W_{ijk}$. As a result, (\ref{relDFF}) represents a relation between the scales $\sqrt{|F|}$ and $M$,
for given gauge coupling $g$. One may however still imagine to tune the real gauge coupling $g$ to achieve any 
desired value of $\sqrt{|F|}$ and $M$ compatibly with this real stationarity condition. Notice finally that after the 
above optimization procedure we are left with $(n-1)(n-2)/2$ free complex $W_{ij}$ and $(n-1)n(n+1)/6 - 1$ 
free complex $W_{ijk}$. This is more than enough to be able to decouple all the $n-2$ complex scalar fields that 
occur in addition to the Goldstino and the Goldstone partners. The simplest possibility is to take the left-over 
$W_{ij}$ to be large and the left-over $W_{ijk}$ to be moderate, so that all these extra scalars become very 
massive and do not induce any sizable negative level-repulsion effect on the masses of the Goldstino and Goldstone 
partners. This shows that the bound (\ref{bound}) can indeed always be saturated by a last tuning of the superpotential.

\section{Renormalizable gauge theories}
\setcounter{equation}{0}

Let us illustrate the implications of our result in the simplest case of renormalizable gauge theories 
with a single $U(1)$ gauge group, where the K\"ahler potential is quadratic and the Killing vector is 
linear:
\bea
\a\a K = \delta_{i \bar \jmath} \Phi^i \bar \Phi^{\bar \jmath} \,,\;\;
X^i = - i q_i \Phi^i \,.
\eea
In this situation, $Q_{i \bar \jmath} = q_i \delta_{ij}$. Moreover, one finds 
$K_i = \delta_{i \bar \jmath} \bar \phi^{\bar \jmath}$ and $K^i = \phi^i$. It then follows that 
$X^i = - i\, Q{}^i{}_j K^j $. Thanks to this last property, and calling $Q^{\text{--}1}{}^i{}_j$ the inverse 
of $Q{}^i{}_j$ restricted to the subspace of non-vanishing charges, one may write:
\bea
\a\a D 
= g\, Q^{\text{--}1}_{i \bar \jmath} X^i \bar X^{\bar \jmath} \,, \\[1mm]
\a\a M^2 
= 2\,g^2 \delta_{i \bar \jmath} X^i \bar X^{\bar \jmath} \,.
\eea

In this simple situation, the scale of the $D$ auxiliary field is related in a very simple and direct way to 
the mass scale $M$. Indeed, it follows from the above definitions that 
$D = \s{\frac 12} g^{\text{--}1} Q^{\text{--}1}_{x \bar x} M^2$. Moreover, the condition (\ref{relDFF}) 
holding at stationary points reads in this case $Q_{f \bar f} |F|^2 = \frac 12 g^{\text{--}1}M^2 D$. 
Using the above relation for $D$, and assuming that $Q_{f \bar f} \neq 0$, this further 
implies that $|F|^2 = \s{\frac 14} g^{\text{--}2} Q^{\text{--}1}_{x \bar x}(Q_{f \bar f})^{\text{--}1} M^4$. 
From these relations, we see that stationary points are possible only if 
\bea
\a\a Q^{\text{--}1}_{x \bar x} Q_{f \bar f} \ge 0 \,.
\label{signs}
\eea
Moreover, the values of the overall $|F|$ and of $|D|$ are related to $M$ and their ratio is fixed
in terms of the values of $Q{}^i{}_j$  along the directions $f^i$ and $x^i$:
\bea
\a\a |D| = \s{\frac 12} g^{\text{--}1} |Q^{\text{--}1}_{x \bar x}| M^2 \,,
\label{relDM} \\[1mm]
\a\a |F| = \s{\frac 12} g^{\text{--}1} \! \raisebox{0pt}{$\sqrt{Q^{\text{--}1}_{x \bar x}(Q_{f \bar f})^{\text{--}1}}$} M^2 \,. 
\label{relFM} \\
\a\a \left|\frac {D}F \right| = \raisebox{0pt}{$\sqrt{Q^{\text{--}1}_{x \bar x} Q_{f \bar f}}$} \,.
\eea
When instead $Q_{f \bar f} = 0$, eq.~(\ref{relDFF}) implies that $|D| = 0$, whereas $|F|$ and 
$M$ can be arbitrary. This is the only situation where $M$ can be adjusted independently of $|F|$.

Notice that we may write down the following simple bound on the relative importance of $D$-type and $F$-type 
supersymmetry breaking, in terms of the pair of charges $q_{\rm min}$ and $q_{\rm max}$ which possess 
the largest possible ratio with the constraint that they have the same sign \cite{DF}:
\bea
\a\a \left|\frac {D}F \right| \le \sqrt{\left|\frac {q_{\rm max}}{q_{\rm min}} \right|} \,.
\eea
This bound can be saturated by choosing the directions $f^i$ and $x^i$ to be the eigenvectors of 
$Q{}^i{}_j$ corresponding to the eigenvalues $q_{\rm max}$ and $q_{\rm min}$.

The scalar masses (\ref{mff}), (\ref{mxaxb}) and (\ref{mfxb}) undergo two relevant simplifications. 
The first is that all the curvature terms drop, since in this case the scalar manifold is flat. 
The second is that due to the relation (\ref{relFM}) the supersymmetric term in $m^2_{x \bar x}$ is forced to be 
of the same order of magnitude as the non-supersymmetric terms. One then finds the following simple expressions:
\bea
\a\a m^2_{f \bar f} = \Big[Q_{x \bar x}^{\text{--}1} \, Q_{f \bar f} \Big] M^2 , \\[1mm]
\a\a m^2_{x \bar x} = \s{\frac 12} \Big[1+ Q_{x \bar x}^{\text{--}1} \, Q_{\hspace{1pt} x \bar x} 
+ Q_{x \bar x}^{\text{--}1}  (Q_{f \bar f})^{\text{--}1} Q^2_{f \bar f} \Big] M^2 , \\[1mm]
\a\a m^2_{f \bar x} = \Big[Q^{\text{--}1}_{x \bar x} \, Q_{f \bar x} \Big] M^2 .
\eea
We observe now that by the restriction (\ref{signs}) and some simple linear algebra, we can get 
some useful constraints on the various pieces of these masses. In particular, we have 
that $Q_{x \bar x}^{\text{--}1} \, Q_{f \bar f} \ge 0$ and 
$Q_{x \bar x}^{\text{--}1}  (Q_{f \bar f})^{\text{--}1} Q^2_{f \bar f} \ge Q_{x \bar x}^{\text{--}1} \, Q_{f \bar f} \ge 0$,
since $Q^2_{f \bar f} \ge (Q_{f \bar f})^2$.
Moreover, $Q_{x \bar x}^{\text{--}1} \, Q_{\hspace{1pt} x \bar x}$ has indefinite sign but becomes 
equal to $1$ whenever $x^i$ is an eigenvector of $Q{}^i{}_j$, and $Q_{x \bar x}^{\text{--}1} \, Q_{\hspace{1pt} f \bar x}$
has indefinite sign but becomes equal to $0$ whenever either $f^i$ or $x^i$ is an eigenvector of $Q{}^i{}_j$.

In this class of models, the masses $m^2_{f \bar f}$, $m^2_{x \bar x}$ and $m^2_{f \bar x}$ depend on the vacuum
point only through the orientation of the direction $x^i$ and the size of $M$. Moreover, by varying the 
vacuum point at fixed $M$ one may achieve all the possible orientations for $x^i$, thanks to the simple 
linear form of $X^i$ and quadratic form of $K$. The optimization of the superpotential defining the bound 
(\ref{bound}) then amounts in this case to optimizing the orientation of the directions $f^i$ and $x^i$, with 
the only constraint that they should be orthogonal. There is then a natural guess for the optimal choice of $f^i$ and $x^i$. 
This consists in choosing these two orthogonal directions to be the eigenvectors of $Q{}^i{}_j$ with largest and 
smallest eigenvalues with common sign, namely $q_{\rm max}$ and $q_{\rm min}$. With such a choice,
$m^2_{f \bar f}$ is maximal, $m^2_{f \bar x}$ vanishes and $2\, m^2_{x \bar x}$ is larger than $m^2_{f \bar f}$. 
The precise values are
\bea
\a\a m^2_{f \bar f} \to \bigg|\frac {q_{\rm max}}{q_{\rm min}} \bigg| M^2 \,,\;\;
m^2_{x \bar x} \to \bigg[1 + \s{\frac 12} \left|\frac {q_{\rm max}}{q_{\rm min}} \right| \bigg] M^2 \,,\;\;
m^2_{f \bar x} \to 0 \,.
\eea
With this choice, one gets that $m^2_{(1)}$, $m^2_{(2)}$ and $m^2_{(3)}$ all coincide with the maximal 
possible value of $m^2_{f \bar f}$. This value certainly represents the maximal possible value for 
$m^2_{(1)}$ taken on its own. But then it must necessarily represent also the maximal possible value 
for $m^2_{(2)}$ and $m^2_{(3)}$, because by construction one has $m^2_{(1)} \ge m^2_{(2)} \ge m^2_{(3)}$ 
for any choice of $f^i$ and $x^i$. This proves that the above choice for $f^i$ and $x^i$ is indeed the optimal 
one, and the bound (\ref{bound}) thus reads in this case
\bea
m^2 = \bigg|\frac {q_{\rm max}}{q_{\rm min}} \bigg| M^2 \,.
\label{boundflat}
\eea
Notice finally that the optimal configuration corresponds in this case to the one that maximizes the size of 
the $D$ auxiliary field relative to the $F$ auxiliary fields: 
\bea
\left|\frac DF \right| \to \sqrt{\bigg|\frac {q_{\rm max}}{q_{\rm min}} \bigg|} \,.
\eea

It is worth emphasizing that in models where the superpotential is adjusted to reach the optimal situation 
described in the previous paragraph, the supersymmetry and gauge-symmetry breaking scales are necessarily 
comparable, so that $F \sim g^{\text{--}1} M^2$. In such a case the vector multiplet plays an important role in the 
low-energy dynamics and gives a sizable contribution to supersymmetry breaking, with $D \sim g^{\text{--}1} M^2 \sim F$. 
The average mass of the Goldstino partners can then be as big as $m^2 \sim g D \sim M^2$. On the other hand, in 
models where the superpotential is instead such that there is a hierarchy between the supersymmetry and 
gauge-symmetry breaking scales, so that $F \ll g^{\text{--}1} M^2$, the situation can never be optimal. 
In such a case the vector multiplet has a small impact on the low-energy dynamics and gives a small contribution 
to supersymmetry breaking, with $D \sim g F^2/M^2 \ll F$. The average mass of the Goldstino partners can then 
still be non-zero and positive, but it is necessarily much smaller than the above-mentioned maximal value: 
$m^2 \sim g D \sim g^2 F^2/M^2 \ll M^2$. This is what happens for instance in models where the gauge symmetry 
is broken by large expectation values for scalars along an almost flat direction, like for example the one 
described in \cite{ADGR}. In this kind of models the heavy vector multiplet can actually be integrated out 
in a supersymmetric way, and the fact that the most dangerous mode is related just to the Goldstino direction 
is then obvious from the beginning. A concrete example of this sort, with a mass spectrum that indeed 
displays the above features, is discussed in some detail in \cite{NRZ}.

Summarizing, we see that in the case of a flat scalar manifold and a linear isometry, the lightest 
scalar field is identified with a partner of the Goldstino, and its square mass is positive. In this particular case, 
one would thus have obtained the same bound by looking only at the Goldstino partners and 
maximizing the smallest of their masses by making the effect of the gauging as large as possible. 
This is however an accidental feature of these models, which is due to the flatness and maximal 
symmetry of the space, as well as the fact that there is a single generator. In next section we will show 
that in the case of curved scalar manifolds, the situation is no-longer so trivial.

\section{Non-linear gauged sigma models}
\setcounter{equation}{0}

Let us next consider the more general case of effective theories with a non-trivial K\"ahler potential
and a single $U(1)$ gauge symmetry generated by a Killing vector of unspecified form:
\bea
\a\a K = K(\Phi^i \bar \Phi^{\bar \jmath}) \,,\;\; X^i = X^i (\Phi^i)\,.
\eea
This situation is of course much more complex than the simple particular case considered in previous 
section. Yet one may try to follow the same steps as before. A major difference is hat since the Killing 
vector $X^i$ is not linear and $K$ is not quadratic, $X^i$ and $K^j$ are no longer linearly related 
through $Q{}^i{}_j$. One may however introduce the new quantity $\tilde Q{}^i{}_j = i X^i K_j/(K^m K_m)$,
which allows to write the relation $X^i = - i\, \tilde Q{}^i {}_j K^j$. 
In the case of renormalizable gauge theories with a phase symmetry, $\tilde Q{}^i{}_j$ coincides with $Q{}^i{}_j$ and is constant,
but in the more general situation considered here $\tilde Q{}^i{}_j$ differs from $Q{}^i{}_j$ and is not constant. 
With this notation, and calling $\tilde Q^{\text{--}1}{}^i{}_j$ the inverse of $\tilde Q{}^i{}_j$ in the subspace where it 
does not vanish, one can then write:
\bea
\a\a D = g\,  \tilde Q^{\text{--}1}_{i \bar \jmath} X^i \bar X^{\bar \jmath} \,, \\[1mm]
\a\a M^2 = 2\, g^2 g_{i \bar \jmath} X^i \bar X^{\bar \jmath} \,.
\eea

In this more complicated case, the auxiliary field $D$ is again related to the mass scale $M$, but in a more 
involved and implicit way. Indeed, from the above definitions one deduces that 
$D = \s{\frac 12} g^{\text{--}1} \tilde Q^{\text{--}1}_{x \bar x} M^2$. 
Moreover, the condition (\ref{relDFF}) implies that at a stationary point $Q_{f \bar f} |F|^2 = \frac 12 g^{\text{--}1}M^2 D$. 
Using the above relation for $D$, and assuming that $Q_{f \bar f} \neq 0$, this further implies that 
$|F|^2 = \s{\frac 14} g^{\text{--}2} \tilde Q^{\text{--}1}_{x \bar x} (Q_{f \bar f})^{\text{--}1} M^4$. 
From these relations, we see that stationary points are possible only if 
\bea
\a\a \tilde Q^{\text{--}1}_{x \bar x}  Q_{f \bar f} \ge 0 \,.
\label{signs2}
\eea
The values of the overall $|F|$ and of $|D|$ are again related to $M$ and their ratio takes as before a simple 
form, but now these relations depend not only on $Q{}^i{}_j$ but also on the new quantities $\tilde Q{}^i{}_j$,
taken respectively along the directions $f^i$ and $x^i$:
\bea
\a\a |D| = \s{\frac 12} g^{\text{--}1} \, |\tilde Q^{\text{--}1}_{x \bar x}| M^2 \,,
\label{relDM2} \\[1mm]
\a\a |F| = \s{\frac 12} g^{\text{--}1} \! \raisebox{0pt}{$\sqrt{\tilde Q^{\text{--}1}_{x \bar x}(Q_{f \bar f})^{\text{--}1}}$} M^2 \,. 
\label{relFM2} \\
\a\a \left|\frac {D}F \right| = \raisebox{0pt}{$\sqrt{\tilde Q^{\text{--}1}_{x \bar x} Q_{f \bar f}}$} \,.
\eea
When instead $Q_{f \bar f} = 0$, eq.~(\ref{relDFF}) implies that $|D| = 0$, whereas $|F|$ and $M$ can be arbitrary. 
As before, this is the only situation where $M$ can be adjusted independently of $|F|$.

In this case, the relative importance of $D$-type and $F$-type supersymmetry breaking depends on the vacuum 
point not only through the direction $x^i$ but also through $Q{}^i{}_j$ and $\tilde Q{}^i{}_j$. Finding an explicit and 
quantitative bound on their ratio is then more difficult. See for instance \cite{K} for some attempts. Nevertheless, 
from the above relations one may still infer a simple although somewhat implicit bound that involves the maximal 
eigenvalue $Q_{\rm max}$ of $Q{}^i{}_j$ and the minimal eigenvalue of $\tilde Q_{\rm min}$ of $\tilde Q{}^i{}_j$, 
with the constraint that these should have the same sign:
\bea
\a\a \left|\frac {D}F \right| \le \sqrt{\left|\frac {Q_{\rm max}}{\tilde Q_{\rm min}}\right|} \,.
\eea
In general, this bound can however not be saturated, because $Q{}^i{}_j$ and $\tilde Q{}^i{}_j$ are different 
matrices that cannot be diagonalized simultaneously, and it is therefore not possible to choose the orthogonal 
directions $f^i$ and $x^i$ in such a way to get simultaneously $Q_{f \bar f} = Q_{\rm max}$ and 
$\tilde Q_{x \bar x} = \tilde Q_{\rm min}$. 

The masses (\ref{mff}), (\ref{mxaxb}) and (\ref{mfxb}) can now be computed more explicitly. In this case
there is an additional contribution coming from the curvature. As before, the relation (\ref{relFM2}) allows 
to rewrite the non-supersymmetric pieces in terms of the same scale as the supersymmetric piece. 
One then finds the following expressions:
\bea
\a\a m^2_{f \bar f} = \Big[\!-\! \s{\frac 14} g^{\text{--}2} \hspace{-1pt} M^2 \hspace{-1pt} 
R_{f \bar f f \bar f} \, \tilde Q_{x\bar x}^{\text{--}1}(Q_{f \bar f})^{\text{--}1} \!
+ \tilde Q_{x \bar x}^{\text{--}1} \, Q_{f \bar f} \Big] M^2 , \\[1mm]
\a\a m^2_{x \bar x} = \s{\frac 12} \Big[1
\hspace{-1pt}-\hspace{-1pt} \s{\frac 12} g^{\text{--}2} \hspace{-1pt} M^2 \hspace{-1pt} R_{f \bar f x \bar x}\, 
\tilde Q_{x\bar x}^{\text{--}1}(Q_{f \bar f})^{\text{--}1} \!
+ \tilde Q_{x \bar x}^{\text{--}1} \, Q_{\hspace{1pt} x \bar x} 
+ \tilde Q_{x \bar x}^{\text{--}1}  (Q_{f \bar f})^{\text{--}1} Q^2_{f \bar f} \Big] M^2 , \\[1mm]
\a\a m^2_{f \bar x} = \Big[\!-\! \s{\frac 14} g^{\text{--}2} M^2 R_{f \bar f f \bar x}\, \tilde Q_{x\bar x}^{\text{--}1}(Q_{f \bar f})^{\text{--}1} \!
+ \tilde Q^{\text{--}1}_{x \bar x} \, Q_{f \bar x} \Big] M^2 .
\eea
There are again various restrictions on the ingredients appearing in these expressions. 
Concerning the contractions of $Q_{i \bar \jmath}$ and $\tilde Q_{i \bar \jmath}$, 
the restriction (\ref{signs2}) implies as before useful constraints.
In particular, we have $\tilde Q_{x \bar x}^{\text{--}1} \, Q_{f \bar f} \ge 0$ and 
$\tilde Q_{x \bar x}^{\text{--}1}  (Q_{f \bar f})^{\text{--}1} Q^2_{f \bar f} \ge \tilde Q_{x \bar x}^{\text{--}1} \, Q_{f \bar f} \ge 0$.
Moreover, $\tilde Q_{x \bar x}^{\text{--}1} \, Q_{\hspace{1pt} x \bar x}$ is indefinite and deviates from $1$ 
even when $x$ is an eigenvector of $Q{}^i{}_j$, whereas $\tilde Q_{x \bar x}^{\text{--}1} \, Q_{\hspace{1pt} f \bar x}$
has indefinite sign but becomes as before equal to $0$ whenever either $f^i$ or $x^i$ is an eigenvector of $Q{}^i{}_j$.
Concerning the contractions of $R_{i \bar \jmath k \bar l}$, on the other hand, there does not seem to exist any 
sharp inequality.

In this class of models, the masses $m^2_{f \bar f}$, $m^2_{x \bar x}$ and $m^2_{f \bar x}$ depend on the vacuum
point not only through the orientation of the direction $x^i$ and the size of $M$, but also through the 
values of $R_{i \bar \jmath k \bar l}$, $Q_{i \bar \jmath}$ and $\tilde Q_{i \bar \jmath}$, 
which are in general not constant. Moreover, it is no longer granted that by 
varying the vacuum point at fixed $M$ one may achieve all the possible orientations for $x^i$.
The optimization of the superpotential defining the bound (\ref{bound}) is then a complicated task,
and does not simply amount to optimizing the orientation of the directions $f^i$ and $x^i$. Moreover,
even ignoring this difficulty, finding the optimal choice is more involved also because of the fact that 
generically it emerges from a competition between the terms that depend only on $Q_{i \bar \jmath}$ 
and $\tilde Q_{i \bar \jmath}$ and those that depend also on $R_{i \bar \jmath k \bar l}$, although there 
may be regimes where one or the other of these two contributions dominates. 
As a consequence of this, we were not able to find any general result 
for this type of models based on curved geometries. We however studied in some detail a few particular 
examples in appendix A, based on simple geometries with covariantly constant curvature and simple 
isometries. The only few remarks that can be made in general concern the behavior of the 
various contractions that appear in the masses $m^2_{f \bar f}$, $m^2_{x \bar x}$ and $m^2_{f \bar x}$ 
when the directions $f^i$ and $x^i$ are varied. To get an idea of what may happen, we may treat 
$f^i$ and $x^i$ as arbitrary directions and enforce the constraints that $g_{i \bar \jmath} f^i \bar f^{\bar \jmath} = 1$, 
$g_{i \bar \jmath} x^i \bar x^{\bar \jmath} = 1$ and $g_{i \bar \jmath} f^i \bar x^{\bar \jmath} = 0$
through Lagrange multipliers. Proceeding in this way, one then finds the following results.
When $Q_{f \bar f}$ is extremal $Q_{f \bar x} = 0$, when $\tilde Q^{\text{--}1}_{x \bar x}$ is extremal 
$\tilde Q^{\text{--}1}_{f \bar x} = 0$, when $Q_{f \bar f} \, \tilde Q^{\text{--}1}_{x \bar x}$ is extremal 
$Q_{f \bar x} \, \tilde Q^{\text{--}1}_{x \bar x} + Q_{f \bar f} \, \tilde Q^{\text{--}1}_{f \bar x} = 0$, and 
finally when $R_{f \bar f f \bar f}$ is extremal $R_{f \bar f f \bar x} = 0$. 

Summarizing, we see that in the case of a curved scalar manifold and a generic isometry, the lightest scalar field
is generically identified with a linear combination of Goldstino and Goldstone partners, and its square mass is 
not necessarily positive. 
In this case, one would thus have obtained a too optimistic bound by looking only at the Goldstino partners 
and maximizing the smallest of their mass. Notice finally that the optimal situation does not necessarily
correspond to the one that maximizes the effect of the gauging. 

\section{Conclusions}
\setcounter{equation}{0}

In this work, we have shown that it is possible to derive an absolute upper bound on the mass of the lightest 
scalar field of a theory with spontaneously broken supersymmetry and local gauge symmetries. This can be 
obtained by focusing on the subset of scalar fields corresponding to the partners of the Goldstino fermion and 
the gauge vector bosons, for which the mass is constrained by symmetry arguments. The resulting bound has 
the property that it can be saturated by adjusting the superpotential. Requiring it to be positive is therefore 
a necessary and sufficient condition on the remaining functions specifying the kinetic terms for the existence 
of a metastable supersymmetry-breaking vacuum. We have shown that by including also the Goldstone 
partners one finds in general a stronger bound than by considering just the Goldstino partners, and we 
have illustrated this fact through several explicit examples. 

Our result has interesting implications on the conditions for the existence of metastable supersymmetry-breaking 
vacua in generic supersymmetric theories with local gauge symmetries. Indeed, the region of parameter space 
where tachyons can be avoided is reduced when one considers not only the Goldstino partners but also 
the Goldstone partners, since there are points where the former have positive square mass while the latter or 
linear combinations of the two have negative square mass. We believe that there may in fact exist models where 
the upper bound derived from just the Goldstino partners is positive whereas the upper bound derived by including also 
the Goldstone partners is negative. In such a situation, one would then find an obstruction against the 
existence of metastable supersymmetry-breaking vacua that comes from the Goldstone partners rather 
than from the Goldstino partners. In the light of this possibility, it would be interesting to apply the result
that we have derived to reexamine the conditions for the existence of metastable supersymmetry-breaking 
vacua in theories where the gauging plays a crucial role. One class of models where this could perhaps 
uncover new instabilities is that of theories with extended supersymmetry, and more specifically those 
where the Goldstino partners do not seem to lead necessarily to tachyons. This is for instance the case 
of $N=2$ theories with non-Abelian vector multiplets and/or charged hyper multiplets. 

To conclude, we would like to comment on the generalization of our result to the case of supergravity 
theories. The only technical difficulty to extend our analysis to that case is the fact that the Goldstino 
direction $f^i$ and the Goldstone directions $x_a^i$ are no longer orthogonal, as a consequence of the 
additional gravitational term in the definition of the auxiliary fields. More precisely, one gets
$g_{i \bar \jmath} f^i \bar x_a^{\bar \jmath} = i g^{\text{--}1} m_{3/2} D_a$. As a consequence, the set of 
vectors $f^i$ and $x_a^i$ can no longer be chosen to form an orthonormal set, although it still represents
a complete set of dangerous directions. The restriction of the mass matrix to this subspace is then no 
longer given just by eq.~(\ref{mat}) but by a more complex expression. As a result, the analysis becomes 
technically more complicated. But for the rest one can apply the same strategy we developed in this paper 
for theories with rigid supersymmetry.

\vskip 20pt

\noindent
{\bf \Large Acknowledgements}

\vskip 10pt

\noindent
This work was supported by the Swiss National Science Foundation. 

\appendix

\section{Explicit examples}
\setcounter{equation}{0}

In this appendix, we study in some detail a few concrete examples to illustrate 
our general results. We focus on models with two fields and one gauge symmetry.
In this situation, the Goldstino and Goldstone directions $f^i$ and $x^i$ are rigidly 
tied and can be parametrized with a single angle $\theta$, which we shall define 
in such a way that the mass $M$ is constant. Another simplification that occurs in 
the two-field case is that one simply has $Q^2_{f \bar f} = (Q_{f \bar f})^2 + |Q_{f \bar x}|^2$.
We shall take $\theta \in [0,2 \pi]$, but in all the examples below the behaviors of 
the masses in the four quadrants are related by simple reflections.

As a first simple example, let us discuss the case of quadratic K\"ahler potential and linear 
Killing vector, which corresponds to a flat scalar manifold with a phase isometry defined by 
positive charges:
\bea
\a\a K = \Phi^1 \bar \Phi^1 + \Phi^2 \bar \Phi^2 \,,\;\; X^i = - i \, \big(q_1 \Phi^1,\, q_2 \Phi^2\big) \,.
\label{model1}
\eea
In this case, we can parametrize the vacuum in the following way:
\bea
\a\a \Phi^i = \s{\frac 1{\sqrt{2}}} \, g^{\text{--}1} M  \, 
\big(q_1^{\text{--}1}\cos \theta ,\, q_2^{\text{--}1} \sin \theta \big)\,.
\eea
The Goldstone and Goldstino directions are then given by 
$x^i = - i\,\big(\cos \theta,\, \sin \theta\big)$ and $f^i = - i\, \big(\sin \theta, - \cos \theta \big)$,
and the metric is clearly trivial: $g_{i \bar \jmath} = \delta_{ij}$.
The relations between $|D|$, $|F|$ and $M^2$ are in this case:
\bea
\a\a |D| = \s{\frac 12} \, g^{\text{--}1} \Big(q_1^{\text{--}1} \cos^2 \theta + q_2^{\text{--}1} \sin^2 \theta \Big) M^2 \,, \\
\a\a |F| = \s{\frac 12} \, g^{\text{--}1} (q_1 q_2)^{\text{--}1/2} M^2 \,.
\eea
We then get:
\bea
\a\a \left|\frac {D}F \right| =  \sqrt{\frac {q_2}{q_1}} \cos^2 \theta + \sqrt{\frac {q_1}{q_2}} \sin^2 \theta  \,.
\eea
In this case $R_{i \bar \jmath k \bar l}$ vanishes identically and we therefore get:
\bea
R_{f \bar f f \bar f} = 0 \,,\;\; 
R_{f \bar f x \bar x} = 0 \,,\;\; 
R_{f \bar f f \bar x} = 0 \,.
\eea
The matrix elements of $Q_{i \bar \jmath}$ are instead given simply by:
\bea
\a\a Q_{f \bar f} = q_2 \cos^2\theta + q_1 \sin^2 \theta \,,\\
\a\a Q_{x \bar x} = q_1 \cos^2 \theta + q_2 \sin^2 \theta \,,\\
\a\a Q_{f \bar x} = (q_1 - q_2) \cos \theta \sin \theta \,.
\eea
The elements $m^2_{f \bar f}$, $m^2_{x \bar x}$, $m^2_{f \bar x}$ and the eigenvalues $m^2_\pm$ of the 
mass matrix are equal to $M^2$ times some functions of $\theta$ and $q_1/q_2$. The behavior of 
$m^2_{f \bar f}/M^2$ and $m_-^2/M^2$ as functions of $\theta$ is shown in fig.~1 for some particular 
choice of $q_1/q_2$. More in general, one finds the following behavior. If $q_1 > q_2$, $m^2_{f \bar f}$ 
and $m^2_-$ both reach their maxima for $\theta = \frac {\pi}2$, and at that point $m^2_{f \bar f}/M^2 = q_1/q_2$, 
$m^2_{x \bar x}/M^2 = \frac 12 (2 + q_1/q_2)$ and $m^2_{f \bar x}/M^2 = 0$, so that $m_-^2/M^2 = q_1/q_2$. 
The optimal direction is therefore $\theta = \frac {\pi}2$, and the bound is 
$m^2/M^2 = q_1/q_2$. If instead $q_2 > q_1$, the situation is similar but with $q_1 \leftrightarrow q_2$
and $\theta \leftrightarrow \frac {\pi}2-\theta$.

\begin{figure}[h]
\vskip 20pt
\begin{center}
\includegraphics[width=0.6\textwidth]{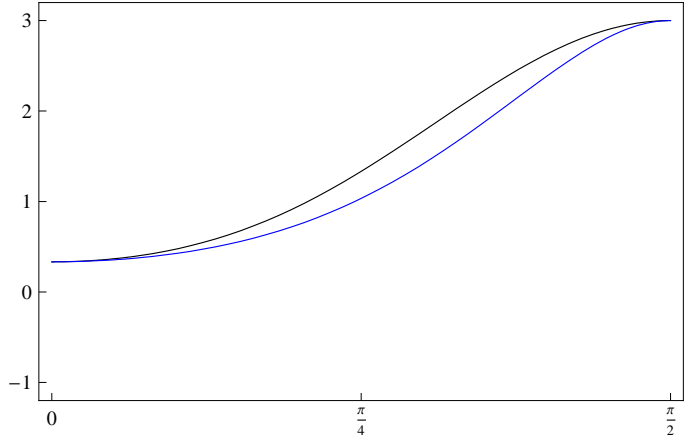}
\caption{Plot of $m^2_{f \bar f}/M^2$ (upper curve) and $m_-^2/M^2$ (lower curve) as functions of $\theta$ 
for the model with quadratic K\"ahler potential and linear Killing vectors defined by (\ref{model1}), with $q_1/q_2 = 3$.}
\end{center}
\end{figure}

As a second simple example, let us discus the case of logarithmic K\"ahler potential and 
constant Killing vector, which corresponds to a constantly and positively curved scalar 
manifold with a shift isometry defined by positive shifts:
\bea
\a\a K = - \Lambda_1^2 \log \bigg(\frac {\Phi^1 + \bar \Phi^1}{\Lambda_1} \bigg)
- \Lambda_2^2 \log \bigg(\frac {\Phi^2 + \bar \Phi^2}{\Lambda_2} \bigg) \,,\;\;
X^i = i \, \big(A_1,\,A_2\big) \,.
\label{model2}
\eea
The two scales $\Lambda_1$ and $\Lambda_2$ define the curvatures of the two field sectors,
whereas the two scales $A_1$ and $A_2$ define the gauge shifts. It is then convenient to 
introduce the following dimensionless parameters:
\bea
\a\a \lambda_1 = \frac {g \Lambda_1}{M} \,,\;\; \lambda_2 = \frac {g \Lambda_2}{M} \,,\;\;
a_1 = \frac {g A_1}{M} \,,\;\; a_2 = \frac {g A_2}{M} \,.
\eea
In this case, we can parametrize the vacuum in the following way, by including 
absolute values to take into account that the fields are in this case restricted to 
have a positive real part:
\bea
\a\a \Phi^i = \s{\frac 1{\sqrt{2}}} \, g^{\text{--}1} M \, 
\big(a_1 \lambda_1 |\sec \theta|,\, a_2 \lambda_2 |\csc \theta| \big) \,.
\eea
The Goldstone and Goldstino directions are then given by 
$x^i = \s{\sqrt{2}} i\,\big(a_1,\, a_2\big)$ and 
$f^i = \s{\sqrt{2}} i\,\big(a_1|\tan \theta|, - a_2 |\cot \theta| \big)$,
whereas $g_{i \bar \jmath} = \frac 12\, {\rm diag} 
\big(a_1^{\text{--}2} \cos^2 \theta,\,a_2^{\text{--}2} \sin^2 \theta\big)$.
The relation between $|D|$, $|F|$ and $M^2$ are in this case:
\bea
\a\a |D| = \s{\frac 1{\sqrt{2}}} \, g^{\text{--}1} \big(\lambda_1 |\cos \theta| + \lambda_2 |\sin \theta| \big) \, M^2 \,, \\
\a\a |F| = \s{\frac 1{\sqrt{2}}} \, g^{\text{--}1} \sqrt{\lambda_1 \lambda_2} \, |2 \cos \theta \sin \theta|^{\text{--}1/2} \, M^2 \,.
\eea
We then get:
\bea
\a\a \bigg|\frac DF \bigg| = \sqrt{|2 \cos \theta \sin \theta|} \,
\bigg(\sqrt{\frac {\lambda_1}{\lambda_2}} |\cos \theta| + \sqrt{\frac {\lambda_2}{\lambda_1}} |\sin \theta| \bigg) \,.
\eea
The contractions of $R_{i \bar \jmath k \bar l}$ are given by
\bea
\a\a R_{f \bar f f \bar f} = 2 g^2 M^{\text{--}2} \big(\lambda_2^{-2} \cos^4 \theta + \lambda_1^{-2} \sin^4 \theta \big) \,,\\
\a\a R_{f \bar f x \bar x} = 2 g^2 M^{\text{--}2} \big(\lambda_2^{-2} + \lambda_1^{-2} \big) \cos^2 \theta \sin^2 \theta \,,\\
\a\a R_{f \bar f f \bar x} = 2 g^2 M^{\text{--}2} \big(\lambda_1^{-2} \sin^2 \theta - \lambda_2^{-2} \cos^2 \theta \big) 
|\cos \theta \sin \theta| \,.
\eea
The matrix elements of $Q_{i \bar \jmath}$ are instead found to be independent of the shifts $a_i$ and 
dominated by the effect of the connection term in their definition, as a result of the fact that the 
Killing vectors are constant:
\bea
\a\a Q_{f \bar f} = \s{\sqrt{2}} \, \big(\lambda_2^{\text{--}1} |\cos \theta| 
+ \lambda_1^{\text{--}1} |\sin \theta| \big) |\cos \theta \sin \theta| \,,\\
\a\a Q_{x \bar x} = \s{\sqrt{2}} \, \big(\lambda_1^{\text{--}1} |\cos^3 \theta| 
+ \lambda_2^{\text{--}1} |\sin^3 \theta| \big) \,,\\
\a\a Q_{f \bar x} = \s{\sqrt{2}} \, \big(\lambda_1^{\text{--}1} |\cos \theta| 
- \lambda_2^{\text{--}1} |\sin \theta| \big) |\cos \theta \sin \theta| \,.
\eea
The elements $m^2_{f \bar f}$, $m^2_{x \bar x}$, $m^2_{f \bar x}$ and the eigenvalues $m^2_\pm$ of the 
mass matrix are equal to $M^2$ times some functions of $\theta$ and $\lambda_1/\lambda_2$.
The behavior of $m^2_{f \bar f}/M^2$ and $m_-^2/M^2$ as functions of $\theta$ is shown in 
fig.~2 for some particular choice of $\lambda_1/\lambda_2$. More in general, one finds the following behavior.
$m^2_{f \bar f}$ reaches its maximum for $\theta = \frac {\pi}4$ and at that point
$m^2_{f \bar f}/M^2 = 1 + \frac 14 (\lambda_1/\lambda_2+\lambda_2/\lambda_1)$, 
$m^2_{x \bar x}/M^2 = 1 + \frac 12 (\lambda_1/\lambda_2+\lambda_2/\lambda_1)$ and 
$m^2_{f \bar x}/M^2 = - \frac 14 (\lambda_1/\lambda_2 - \lambda_2/\lambda_1)$, so that $m_-^2/M^2$ is 
smaller-or-equal than $m^2_{f \bar f}/M^2$. The maximum of $m_-^2/M^2$ occurs instead for some $\theta \le \frac {\pi}4$ if
$\lambda_1 > \lambda_2$ and for some $\theta \ge \frac {\pi}4$ if $\lambda_1 < \lambda_2$, and takes a value that is 
smaller than $1 + \frac 14 (\lambda_1/\lambda_2+\lambda_2/\lambda_1)$. 
For $\lambda_1 \simeq \lambda_2$, the optimal direction is $\theta \simeq \frac {\pi}4$ and the bound is 
$m^2/M^2 \simeq \frac 32$, which is identical to the one that one would have obtained by looking just 
at the Goldstino direction. For $\lambda_1 \gg \lambda_2$, on the other hand, a numerical study shows 
that the optimal direction is $\theta \simeq 0.67$ and the bound is $m^2/M^2 \simeq 0.13 \, \lambda_1/\lambda_2$, 
which is a factor $1.86$ smaller than the one that one would have inferred by looking just at the Goldstino direction, 
although still positive. For $\lambda_1 \ll \lambda_2$, the situation is similar but with $\lambda_1 \leftrightarrow \lambda_2$ 
and $\theta \leftrightarrow \frac {\pi}2-\theta$.

\begin{figure}[h]
\vskip 20pt
\begin{center}
\includegraphics[width=0.6\textwidth]{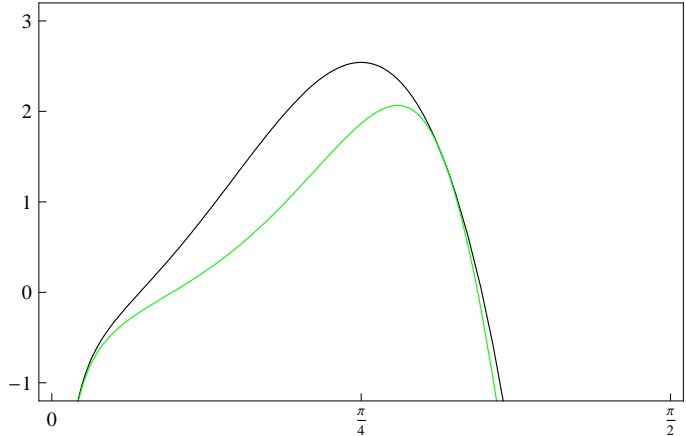}
\caption{Plot of $m^2_{f \bar f}/M^2$ (upper curve) and $m_-^2/M^2$ (lower curve) as functions of $\theta$ for the model 
with logarithmic K\"ahler potential and constant Killing vectors defined by (\ref{model2}), with 
$\lambda_1/\lambda_2= \frac 16$.}
\end{center}
\end{figure}

As a third slightly more complicated and richer example, let us finally discus the case of logarithmic 
K\"ahler potential and linear Killing vector, which corresponds to a constantly and positively curved scalar 
manifold with a phase isometry defined by positive charges:
\bea
\a\a K = - \Lambda_1^2 \log \bigg(\hspace{-1pt}1 \hspace{-1pt}-\hspace{-1pt} 
\frac {\!\Phi^1 \bar \Phi^1\!}{\Lambda_1^2}\hspace{1pt} \bigg)
- \Lambda_2^2 \log \bigg(\hspace{-1pt}1 \hspace{-1pt}-\hspace{-1pt} 
\frac {\!\Phi^2 \bar \Phi^2\!}{\Lambda_2^2}\hspace{1pt} \bigg) \,,\;\;
X^i = - i\, \big(q_1 \Phi^1\hspace{-1pt}, q_2 \Phi^2\big) \,.
\label{model3}
\eea
The two scales $\Lambda_1$ and $\Lambda_2$ define as before the curvatures of the two 
field sectors. It turns out that by varying the overall scale of these curvatures with respect to the 
vector mass scale, this new model interpolates between the two previous ones. This can be 
seen as follows. The small curvature limit corresponds to 
take $\Lambda_i$ large and $\Phi^i$ finite, so that $\Phi^i/\Lambda_i$ is close to $0$. 
In this limit one can keep the same coordinates and just expand the logarithm in $K$. In this 
way one then recovers the model (\ref{model1}). The large curvature limit corresponds 
instead to take $\Lambda_i$ small and $\Phi^i$ also small, so that $\Phi^i/\Lambda_i$ 
is close to $1$. In this limit, it is convenient to change coordinates to describe the model
in a more transparent way. The appropriate reparametrization turns out to be 
$\Phi^i/\Lambda_i \to (1 - \frac 12 \Phi^i/\Lambda_i)/(1 + \frac 12 \Phi^i/\Lambda_i)$.
Discarding an irrelevant K\"ahler transformation, one then finds 
$K \to - \sum_i \Lambda_i^2 \log ((\Phi^i + \bar \Phi^i)/\Lambda_i)$ and 
$X^i \to i \, q_i \Lambda_i \, (1 - \frac 14 \Phi^{i2}/\Lambda_i^2)$.
In these new coordinates, $\Phi^i/\Lambda_i$ is close to $0$. In this limit one then 
manifestly recovers the model (\ref{model2}) with the same field parametrization and shifts 
given by $A_i = q_i \Lambda_i$. To parametrize the effects of the curvatures, we introduce 
as before the dimensionless parameters
\bea
\a\a \lambda_1 = \frac {g \Lambda_1}{M} \,,\;\; \lambda_2 = \frac {g \Lambda_2}{M} \,.
\eea
It will also be useful to introduce the short-hand notation
\bea
u(\theta) = H \bigg(\frac {\cos \theta}{q_1 \lambda_1}\bigg) \,,\;\;
v(\theta) = H \bigg(\frac {\sin \theta}{q_2 \lambda_2}\bigg) \,.
\eea
where $H(x)$ is the following monotonically decreasing function:
\bea
\a\a H(x) = \frac {\sqrt{1+ 2 \, x^2} - 1}{x^2} \simeq
\Bigg\{
\begin{array}{l}
\! 1 \;\;\;\;\;\;\;,\;\; |x| \ll 1 \\[1.5mm]
\! \s{\sqrt{2}}/|x| \,,\;\; |x| \gg 1 \\
\end{array}
\,.
\eea
In this case, we can parametrize the vacuum in the following way:
\bea
\a\a \Phi^i = \s{\frac 1{\sqrt{2}}} \, g^{\text{--}1} M \, \big(q_1^{\text{--}1} u(\theta)  \cos \theta ,\,
q_2^{\text{--}1} v(\theta) \sin \theta \big) \,.
\eea
The Goldstone and Goldstino directions then read  
$x^i = - i \,\big(u(\theta) \cos \theta,\, v(\theta) \sin \theta \big)$ and 
$f^i = - i \, \big(u(\theta) \sin \theta, - v(\theta) \cos \theta \big)$, and the metric 
is $g_{i \bar \jmath} = {\rm diag} \big(1/u^{2}(\theta),\,1/v^{2}(\theta) \big)$.
The relation between $|D|$, $|F|$ and $M^2$ are in this case:
\bea
\a\a |D| =  \s{\frac 12} g^{\text{--}1} \, 
\big(q_1^{\text{--}1} u(\theta)  \cos^2 \theta + q_2^{\text{--}1} v(\theta) \sin^2 \theta\big) M^2 \,, \\
\a\a |F| = \s{\frac 12} g^{\text{--}1} \sqrt{\frac{q_1^{\text{--}1} u(\theta) \cos^2 \theta + q_2^{\text{--}1} v(\theta) \sin^2 \theta}
{q_2 \big[2/v(\theta)-1\big] \cos^2 \theta + q_1 \big[2/u(\theta)-1\big] \sin^2 \theta}} \, M^2 \,.
\eea
We then get:
\bea
\a\a \bigg|\frac DF \bigg| = \sqrt{\sqrt{\frac {q_2}{q_1}} u(\theta)  \cos^2 \theta 
+ \sqrt{\frac {q_1}{q_2}} v(\theta) \sin^2 \theta} \nn \\
\a\a \hspace{35pt} \times \sqrt{\sqrt{\frac {q_2}{q_1}} \big[2/v(\theta)-1\big] \cos^2 \theta 
+ \sqrt{\frac {q_1}{q_2}} \big[2/u(\theta)-1\big] \sin^2 \theta}
\,.
\eea
The contractions of $R_{i \bar \jmath k \bar l}$ are given by
\bea
\a\a R_{f \bar f f \bar f} = 2 g^2 M^{\text{--}2} \big(\lambda_2^{-2} \cos^4 \theta + \lambda_1^{-2} \sin^4 \theta \big) \,, \\
\a\a R_{f \bar f x \bar x} = 2 g^2 M^{\text{--}2} \big(\lambda_2^{-2} + \lambda_1^{-2} \big) \cos^2 \theta \sin^2 \theta \,, \\
\a\a R_{f \bar f f \bar x} = 2 g^2 M^{\text{--}2} \big(\lambda_1^{-2} \sin^2 \theta - \lambda_2^{-2} \cos^2 \theta \big) 
\cos \theta \sin \theta \,.
\eea
The matrix elements of $Q_{i \bar \jmath}$ are instead found to be:
\bea
\a\a Q_{f \bar f} = q_2 \big[2/v(\theta)-1\big] \cos^2 \theta + q_1\big[2/u(\theta)-1\big]\sin^2 \theta \,, \\
\a\a Q_{x \bar x} = q_1 \big[2/u(\theta)-1\big] \cos^2 \theta + q_2 \big[2/v(\theta)-1\big] \sin^2 \theta \,, \\
\a\a Q_{f \bar x} = \big(q_1\big[2/u(\theta)-1\big] - q_2 \big[2/v(\theta)-1\big] \big) \cos \theta \sin \theta \,.
\eea
The elements $m^2_{f \bar f}$, $m^2_{x \bar x}$, $m^2_{f \bar x}$ and the eigenvalues $m^2_\pm$ of the 
mass matrix are equal to $M^2$ times some functions of $\theta$, $\lambda_1/\lambda_2$, $q_1/q_2$ and 
$q_1 q_2 \lambda_1 \lambda_2$. The behavior of $m^2_{f \bar f}/M^2$ and $m_-^2/M^2$ as functions of 
$\theta$ is shown in fig.~3 for some particular choice of $\lambda_1/\lambda_2$, $q_1/q_2$ and 
$q_1 q_2 \lambda_1 \lambda_2$. More in general, one finds the following behavior. $m^2_{f \bar f}$ and 
$m_-^2$ reach maxima for two different values of $\theta$, and the maximal value of $m_-^2$ is always 
smaller than the maximal value of $m^2_{f \bar f}$. This shows once again that the bound that one would 
have inferred by looking only at the Goldstino direction is weaker than the bound $m^2$ that one obtains 
by taking into account also the Goldstone direction. One moreover verifies that in the limit $\lambda_i \gg 1$ 
one recovers the behavior of the model with quadratic $K$ and linear $X^i$ with charges $q_i$, whereas 
in the limit $\lambda_i \ll 1$ one reaches the behavior of the model with logarithmic $K$ and constant $X^i$ 
with shifts $A_i = q_i \Lambda_i$.

\begin{figure}[h]
\vskip 20pt
\begin{center}
\includegraphics[width=0.6\textwidth]{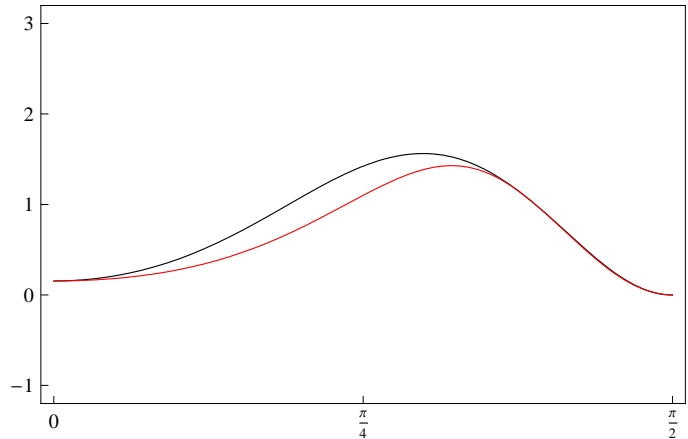}
\caption{Plot of $m^2_{f \bar f}/M^2$ (upper curve) and $m_-^2/M^2$ (lower curve) as functions of $\theta$ for the model 
with logarithmic K\"ahler potential and linear Killing vectors defined by (\ref{model3}),
with $\lambda_1/\lambda_2= \frac 16$, $q_1/q_2 = 3$ and $q_1 q_2 \lambda_1 \lambda_2 = 1$.}
\end{center}
\end{figure}

\end{document}